\documentclass[conference]{IEEEtran}
\usepackage{cite}

\ifCLASSINFOpdf
  \usepackage[pdftex]{graphicx}
\else
\fi
\usepackage{tikz}
\usepackage{amsmath}

\usepackage{algorithmic}
\usepackage{textcomp}
\usepackage{booktabs}
\usepackage[ruled,vlined,linesnumbered]{algorithm2e}
\SetKwComment{Comment}{$\triangleright$\ }{}
\SetKwInOut{Variables}{Variables}
\usepackage{subfigure}
\usepackage{amsmath}
\usepackage{threeparttable}
\usepackage{multirow}
\usepackage{float}
\usepackage{makecell}
\usepackage{balance}
\usepackage{colortbl}
\usepackage{listings}
\usepackage{makecell}
\usepackage{marvosym}
\usepackage[hidelinks]{hyperref}
\definecolor{codegreen}{rgb}{0,0.6,0}
\definecolor{codegray}{rgb}{0.5,0.5,0.5}
\definecolor{codepurple}{rgb}{0.58,0,0.82}
\definecolor{backcolour}{rgb}{0.95,0.95,0.92}

\newcommand{\ie}{\textit{i.e., }}
\newcommand{\eg}{\textit{e.g., }}
\newcommand{\tool}{\textsc{ShapFuzz}}

\definecolor{mygreen}{rgb}{0,0.6,0}
\definecolor{mygray}{rgb}{0.5,0.5,0.5}
\definecolor{mymauve}{rgb}{0.58,0,0.82}
\begin{document}
%
\title{\tool{}: Efficient Fuzzing via Shapley-Guided Byte Selection}









%
\author{\IEEEauthorblockN{Kunpeng Zhang\IEEEauthorrefmark{1}$^\star$,
Xiaogang Zhu\IEEEauthorrefmark{2}$^\star$,
Xi Xiao\IEEEauthorrefmark{1}\textsuperscript{\Letter}, 
Minhui Xue\IEEEauthorrefmark{3},
Chao Zhang\IEEEauthorrefmark{4}\IEEEauthorrefmark{5},
Sheng Wen\IEEEauthorrefmark{2}
}

\IEEEauthorblockA{\IEEEauthorrefmark{1}Shenzhen International Graduate School, Tsinghua University}
\IEEEauthorblockA{\IEEEauthorrefmark{2}Swinburne University of Technology}
\IEEEauthorblockA{\IEEEauthorrefmark{3}CSIRO's Data61\\ \IEEEauthorrefmark{4}Tsinghua University\\
\IEEEauthorrefmark{5}Zhongguancun Laboratory}
}


\IEEEoverridecommandlockouts
\makeatletter\def\@IEEEpubidpullup{6.5\baselineskip}\makeatother
\IEEEpubid{\parbox{\columnwidth}{
    Network and Distributed System Security (NDSS) Symposium 2024\\
    26 February - 1 March 2024, San Diego, CA, USA\\
    ISBN 1-891562-93-2\\
    https://dx.doi.org/10.14722/ndss.2024.23134\\
    www.ndss-symposium.org
}
\hspace{\columnsep}\makebox[\columnwidth]{}}

\maketitle

\begin{abstract}
Mutation-based fuzzing is a popular and effective method for bug exposure and discovery of unseen code in programs. 
However, only a few studies have focused on quantifying the importance of input bytes. The importance of each input byte is determined by its contribution degree in discovering new code.
Previous work often focused on obtaining the relationship between input bytes and path constraints, ignoring the fact that not all constraint-related bytes can discover new code.
In this paper, we conduct Shapley analysis to understand the effect of byte positions on fuzzing performance, and find that some byte positions contribute more than others and this property often holds across different seeds.
Based on this observation, we propose a novel solution, called \tool{}, to guide byte selection and mutation in fuzzing processes.
Specifically, \tool{} updates Shapley values (importance) of bytes when each input is tested during fuzzing with a low overhead. 
It utilizes contextual multi-armed bandit algorithm to make a trade off between mutating high Shapley value bytes and low-frequently chosen bytes.
We implement a prototype of this solution based on AFL++, \ie  \tool{},
and evaluate it against ten state-of-the-art fuzzers, including five byte-scheduling fuzzers and five commonly used fuzzers.
Compared to byte-scheduling fuzzers, \tool{} discovers more edges. It also exposes more bugs than the best baseline on three different sets of initial seeds.
\tool{} exposes 20 more bugs than the best commonly used fuzzers, and discovers 6 more CVEs than the baseline on MAGMA.
Furthermore, \tool{} discovers 11 new bugs on the latest versions of 6 widely used programs, and 3 bugs of them are confirmed by vendors.
{\let\thefootnote\relax\footnote{{$\star$~}Kunpeng Zhang and Xiaogang Zhu contribute equally to this paper.}
\let\thefootnote\relax\footnote{{\textsuperscript{\Letter}~}Corresponding author.}

}



\end{abstract}


%

\section{Introduction}

Mutation-based fuzzing is widely used to detect software vulnerabilities among both academic and industrial practitioners~\cite{zhu2022roadmap}.
In general, to mutate an input, there are two core choices to make: which bytes to mutate (byte selection) and what values to use (value update).
To mutate bytes, state-of-the-art solutions often rely on extra analysis to recognize the bytes related to path constraints.
The first category of solutions utilizes taint analysis to track constraint-related bytes. TaintScope~\cite{taintscope}, Dowser~\cite{dowser} and Angora~\cite{angora} are examples of the first category.
The second category indirectly infers the dependence between bytes and constraints. Members of this category include GREYONE~\cite{greyone}, REDQUEEN~\cite{redqueen}, ProFuzzer~\cite{profuzzer} and PATA~\cite{pata}.
The third category finds where to mutate by extracting information from historical data. Examples are NEUZZ~\cite{neuzz}, MTFuzz~\cite{mtfuzz} and EMS~\cite{ems}. 

While mutating constraint-related bytes indeed improves the efficiency of code discovery, not all constraint-related bytes are able to discover new code. In fact, our experiments show that only 18\% of constraint-related bytes can be mutated to successfully discover new areas (Section~\ref{subsec:cmp_infer}). 
For some target programs, this ratio is even less than 3\%.
However, existing solutions treat all constraint-related bytes equally, wasting time and energy on the ones that cannot discover new code.
Moreover, the extra analysis stage required by existing solutions is often time-consuming due to its byte-by-byte investigation~\cite{greyone, profuzzer, pata}.
For deep learning-based fuzzers~\cite{neuzz, mtfuzz}, the extra stage may even fail the process of fuzzing because a large input size will cause out-of-memory errors when building models.

To solve these problems, our key idea is to quantify the \textit{importance of constraint-related bytes} (\ie the degree to which a byte contributes to the discovery of new code), without going through the extra analysis stage.
Basically, the discovery of new edges is due to the cooperation of certain bytes, or more precisely, their cooperative mutations. This motivates us to take the perspective of Cooperative Game Theory~\cite{Shapley} and study the byte scheduling part of fuzzing as a process of Shapley analysis~\cite{shap}.
Shapley analysis is widely used to quantify the contribution of each player to the result.
By quantifying the importance of input bytes, we can find the bytes that are more likely to discover new code and do more mutations on those bytes. This solution does not require going through an extra analysis stage to obtain the importance because the Shapley analysis can be integrated into the process of mutation.
 
We have conducted a set of fuzzing experiments and performed Shapley analysis to determine the contribution of each byte to discovering new code. 
Interestingly, the analysis demonstrated that (1) \textit{the discovery of new code largely depends on a small portion of byte positions}, and (2) \textit{the mutation of the same byte positions can repeatedly discover new code}.
We observe that these two findings match the nature of program logic. Specifically, some input bytes are repeatedly utilized in the path constraints that prevent fuzzing from discovering code. Our experiments also demonstrated that 86\% of constraint-related bytes are related to multiple CMP instructions used in path constraints (Section~\ref{sec:motivation}). 
In other words, repetitive mutation of high-importance bytes (the small portion) can efficiently solve constraints in different paths.

While Shapley analysis sounds promising in guiding byte mutations, the full Shapley analysis of a byte requires that all the possible inputs related to that byte to be generated and tested. 
Then, no new inputs can be generated for the byte after the full Shapley analysis, making the Shapley values useless for guiding fuzzing. 
Therefore, we propose to utilize \textit{temporary} Shapley values calculated by \textit{incomplete} Shapley analysis to guide fuzzing.
The aforementioned empirical observation implies that this is feasible.
The bytes that temporarily show high Shapley values are still likely related to multiple unseen path constraints.

In this paper, we propose \tool{} to boost the efficiency of byte selection based on Shapley analysis. Specifically, we formulate the solution to byte selection as a Shapley analysis across multiple seed inputs.
To reduce the overhead of calculating Shapley values, we classify seeds into different \textit{families} and share Shapley values in each family, and further transform the Shapley value calculation into a form that can be gradually updated by each seed input. 
We further utilize contextual multi-armed bandit (CMAB) to maximize the efficiency of fuzzing when we use temporary Shapley values.
On the one hand, high Shapley value bytes might not be useful in some situations, \eg when all related constraints of a byte have been explored. 
On the other hand, to update the temporary Shapley values, we need to provide opportunities to low-frequently chosen bytes. Therefore, CMAB makes a trade-off between mutating high Shapley value bytes and low-frequently chosen ones.

We implement a prototype \tool{} based on AFL++~\cite{aflpp}.
We evaluate it on two third-party benchmarks, including 1) UNIFUZZ~\cite{unifuzz}, which contains real-world programs, and 2) MAGMA~\cite{magma}, which provides a series of real-world programs with known bugs. 
We compare their performance with 10 widely-used fuzzers, including 5 byte-scheduling fuzzers (GreyOne~\cite{greyone}, ProFuzzer~\cite{profuzzer}, Angora~\cite{angora}, PreFuzz~\cite{prefuzz} and NEUZZ~\cite{neuzz}) and 5 commonly used fuzzers (AFL++~\cite{aflpp}, MOPT~\cite{MOPT}, AFLFast~\cite{aflfast}, FairFuzz~\cite{fairfuzz} and AFL~\cite{afl}).
Compared with byte-scheduling fuzzers, when all initial seeds are given, \tool{} discovers 4170 more edges and exposes 19 more bugs than the best baseline.
In addition, when given two initial seed sets with lengths less than 10,000 and 1,000 bytes, \tool{} is still able to discover the majority of edges and expose most of the bugs.
\tool{} discovers 20 to 68 more bugs than what the five commonly used fuzzers do, and exposes six more CVEs than what the best of the five fuzzers does.
Furthermore, \tool{} discovers 11 new bugs, 3 of which are confirmed by vendors.

We summarize our main contributions as follows.
\begin{itemize}
    \item Based on Shapley analysis, we perform empirical experiments to study the contributions of bytes in terms of code discovery. The results indicate that the repeated mutation of a small portion of positions can improve the efficiency of code discovery. We ascertain that the main derivative of the Shapley results is that a byte may be related to multiple path constraints.
  
    \item We formalize byte selection as Shapley analysis and transform the calculation of Shapley values into a form that can be updated gradually during fuzzing. We propose \tool{}, which uses Shapley values to guide the byte selection process. It presents a contextual multi-armed bandit approach to optimize Shapley-guided byte selection.
    
    
    \item We implement \tool{} based on AFL++, and evaluate it on UNIFUZZ and MAGMA platforms. The results indicate that \tool{} outperforms various state-of-the-art fuzzers in terms of edge coverage and bug discovery. We open source \tool{} at \url{https://github.com/ShapFuzz/ShapFuzz}.
\end{itemize}

\section{Background and Motivation} \label{sec:motivation}

The aim of coverage-guided fuzzing is to expand the code coverage, which motivates us to study the contributions of bytes in inputs in terms of discovering new coverage.
In this section, we study the contributions (or importance) of bytes  in terms of code discovery.

\subsection{Shapley Analysis and CMAB}

In this section, we introduce the concept of Shapley analysis and contextual multi-armed bandit.

\subsubsection{Shapley Analysis} 
In cooperative game theory, a coalition of players participate in a game and obtain a gain.
The key question is how much each player contributes to the coalition.
Many methods have been proposed to address this problem.
The best-known solution to this question is the Shapley value~\cite{Shapley}, a method for distributing gain among players in a cooperative game.
Specifically, the Shapley value is the weighted average of each player's marginal contributions to all possible coalitions of players.
Moreover, the Shapley value has been applied in numerous fields, such as quantifying the importance of a feature in deep learning models~\cite{fryer2021shapley}.

\subsubsection{CMAB} 
The multi-armed bandit problem is a model to study the trade-offs between exploring new knowledge and exploiting the current reliable knowledge~\cite{lai1985asymptotically}. 
Formally, a bandit problem consists of a set of $N$ arms. 
At the trial $t$, the player pulls an arm $i$ and receives a reward $r_{t}$.
The goal of the player is to maximize the total reward in finite trials.
However, the player can estimate the expected reward for each arm only through repeated trials.
Thus, the player needs to make a trade-off between finding out the reward expectation of each arm and choosing the arm with the highest reward.
The contextual bandit problem is a variant of the multi-armed bandit problem~\cite{auer2002finite}.
In each trial, the player is presented with side information (\ie context) before making decision.
Since the expected reward of arms may be different in different contexts, the player needs to consider the influence of context when choosing an arm.

\begin{figure}[!t]
\centering
\includegraphics[width=0.9\linewidth]{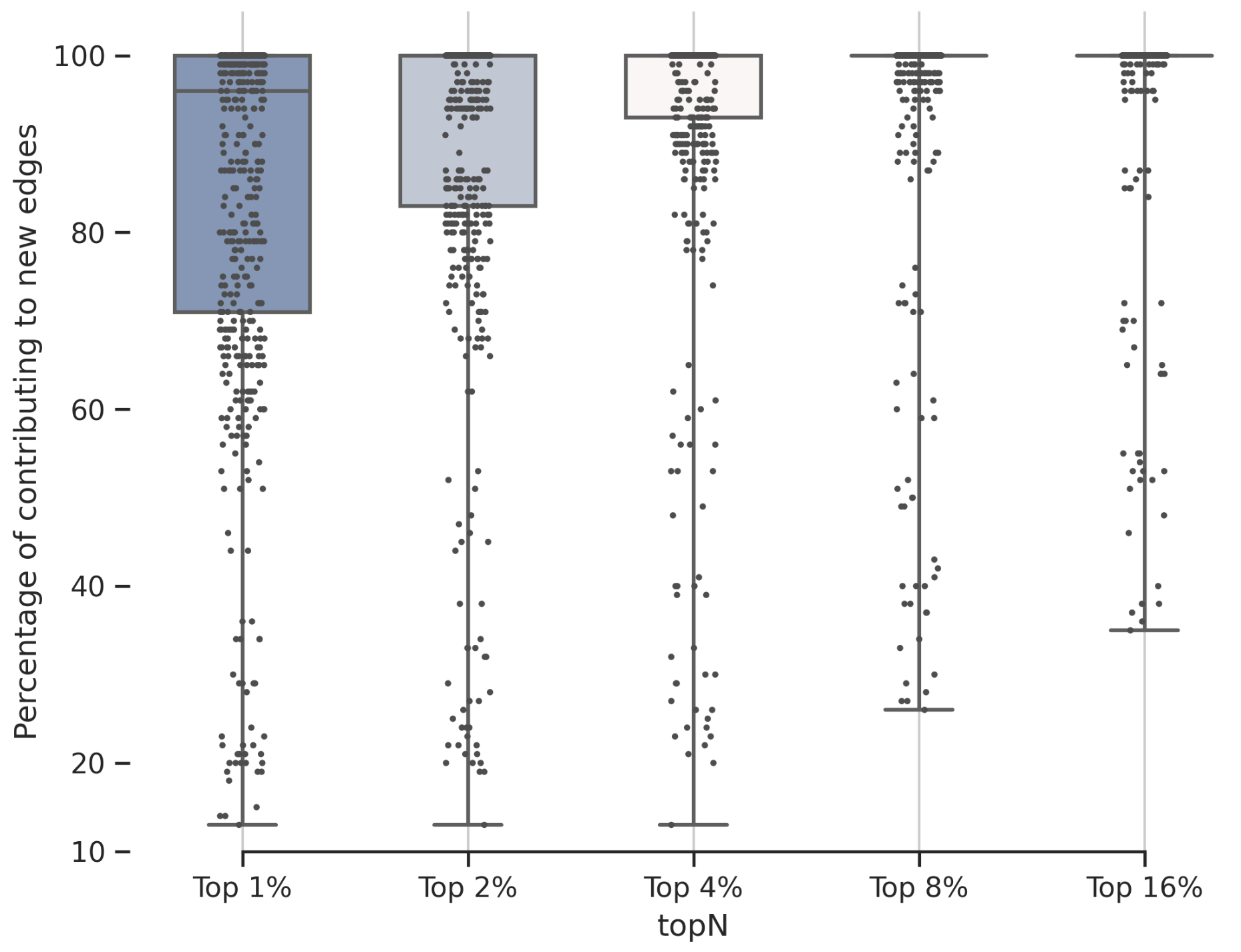}
\caption{The boxplot of contributions of bytes. Y\% of new edges are contributed by the top X\% of bytes. The top X\% of bytes refers to the first X\% bytes with the largest Shapley values. Each dot is a seed of a program.
}
\label{shap_exp_per}
\end{figure}

\subsection{Motivation}
Since bytes are collaboratively mutated to explore code, we use the concept of Shapley value, which is a solution concept in Cooperative Game Theory, to analyze the contributions of bytes. The core idea of Shapley value is to traverse all possible combinations and calculate the average marginal contribution of a byte based on the gains of those combinations. More details are explained in Section~\ref{subsec:shapval}.

\subsubsection{Experiments of Shapley Analysis}  
To calculate Shapley values of bytes in a seed, we regard \textit{the number of new edges} discovered by a combination as the gain. However, a seed with length $N$ has $256^N$ possible combinations, which are too many to be tested. To obtain a relatively accurate Shapley value for a byte, we run fuzzing with random mutation for 48 hours for \textit{a single seed}. 
Specifically, we use AFL++ to run fuzzing on 18 programs with a set of initial seeds for 48 hours. Each fuzzing experiment is only performed on a single initial seed of a program and we do not update the seed queue but continuously mutate the selected initial seed.
The experiments are repeated six times to avoid the influence of randomness. The 18 programs, such as \texttt{nm}, \texttt{tiff2bw}, \texttt{flvmeta}, \texttt{imginfo}, \texttt{infotocap}, \texttt{lame}, \texttt{mp42aac} and \texttt{mp3gain}, include 9 different types of inputs that are \texttt{elf}, \texttt{tiff}, \texttt{pdf}, \texttt{text}, \texttt{mp4}, \texttt{flv}, \texttt{wav}, \texttt{jpg} and \texttt{mp3}. During fuzzing, if a mutation discovers new edges, we will recover all the subsets of related bytes and test the result. For example, if the mutation of positions $\{1,3,4\}$ discovers new edges (the gain $R$), all the subsets $\{1\}/\{3\}/\{4\}/\{1,3\}/\{1,4\}/\{3,4\}$ will be recovered, and their results will be tested. This ensures that all possible combinations related to the gain $R$ are analyzed.
Finally, based on these results, we can calculate the Shapley values for positions $1$, $3$, and $4$.
Notably, we remove the mutators that can change the length of a seed so that each byte does not change its relative position.

\begin{figure}[!t]
\centering
\includegraphics[width=0.9\linewidth]{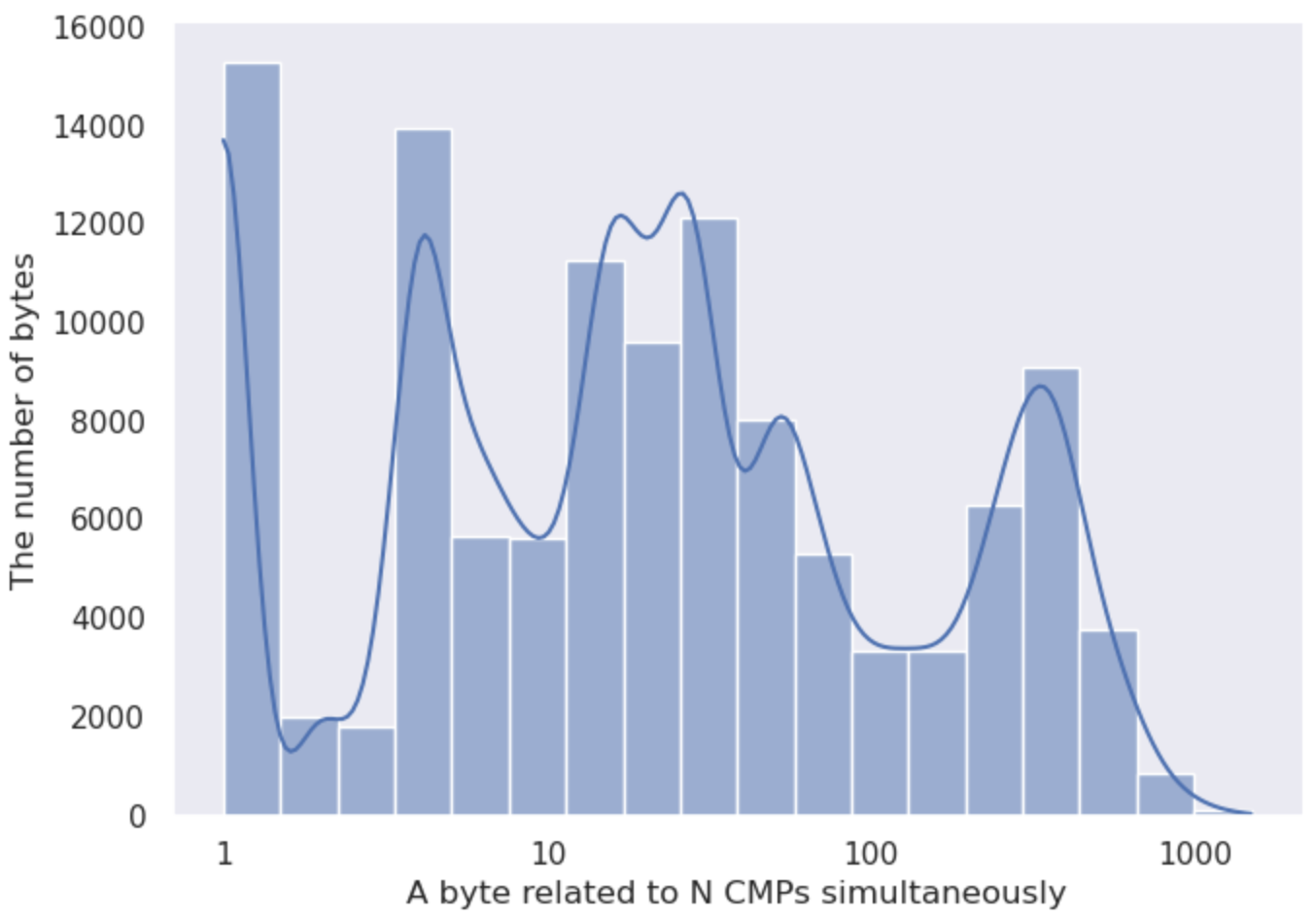}
\caption{Statistics of the relationship between bytes and CMPs across 16 programs.}
\label{cmp}
\end{figure}

\begin{figure*}[t!]
\centering
\includegraphics[width=0.9\linewidth]{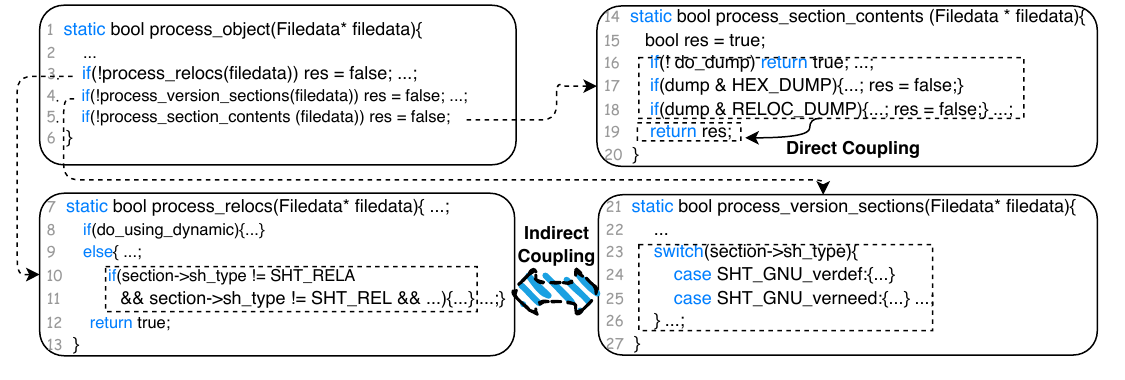}
\caption{The reason for the results of SHAP analysis. The code is extracted from \texttt{readelf}. 
}
\label{example_readelf}
\vspace{-2mm}
\end{figure*}

The experiment results are shown in Figure~\ref{shap_exp_per}, where each dot means that for a program and an initial seed, Y\% of new edges are contributed by top X\% bytes. The results indicate that a small portion of bytes contribute the most to discovering new edges.
For example, the top 4\% positions contribute to the discovery of 97.43\%, 88.72\%, 86.10\%, 94.16\%, 87.96\%, and 100.00\% edges on \texttt{pdftotext}, \texttt{tcpdump}, \texttt{infotocap}, \texttt{strip}, \texttt{objdump}, and \texttt{flvmeta}, respectively.
On some programs, such as \texttt{tiffsplit}, the top 1\% bytes can contribute more than 90\% of new edges.
This also indicates that some bytes repeatedly discover new code during fuzzing.

\subsubsection{Interpretability of the Shapley Results}
In a program input, not all bytes pertain to path constraints. Even for the bytes related to the same path constraint, they contribute differently to solving the constraint. For instance, in the snippet $if ((int)a > 0xeffff)$, the mutation of the two high bytes is more likely to satisfy the condition while the mutation of the two low bytes may not be able to satisfy it. 
Moreover, a byte may be related to more than one path constraint, which increases the Shapley value of the byte. From the perspective of program logic, this is the result of path coupling. 
On the one hand, along the same execution path, the flipping of a path constraint has to consider the results of other path constraints. For example, in Figure~\ref{example_readelf}, the path constraint in line 5 is defined by the function \texttt{process\_section\_contents()}, whose boolean output is determined by the constraints of lines 16, 17 and 18. Thus, input bytes related to lines 16, 17 and 18 also contribute to the solving of constraint in line 5. On the other hand, different program logics may process the same bytes of inputs. For instance, in Figure~\ref{example_readelf}, both the constraints of lines 10 and 23 check the value of \texttt{sh\_type}, which represents the types of sections in the program.

To further investigate the generality of \textit{a byte being related to multiple path constraints simultaneously}, we take the CMP operator as an example and analyze the overlapping situations of related bytes among CMPs in 16 programs, including \texttt{nm}, \texttt{tiff2bw}, and so on.
We use the GreyOne's FTI~\cite{greyone} method to analyze the related bytes of CMP, which has the characteristic of no false positives. Specifically, given a CMP and a seed, if changing byte $i$ of the seed leads to a change in the value of CMP $j$, then byte $i$ and CMP $j$ are related.
Finally, we combine the analysis results of all programs to obtain the final experimental results, as shown in Figure~\ref{cmp}.
We found that most bytes are related to multiple CMPs. For instance, 86\% of bytes are related to more than two CMPs. There are 6,000 bytes that are related to 10 CMPs at the same time. This experiment shows that the sharing of related bytes between CMPs is a common phenomenon.

\subsubsection{Insight}
Since only a small portion of bytes contribute the most to the discovery of new code, \textit{the Shapley analysis can be utilized to obtain those high-importance bytes and more energy is assigned to them during fuzzing}. 
Because it is common that a constraint-related byte is related to multiple CMP instructions, focusing on the high-importance bytes can improve the efficiency of code discovery in unseen paths. Although this insight has been validated in the programs we tested, it is essential to acknowledge that additional validation on a broader range of programs is necessary to establish its generalizability.

\section{Design of \tool} \label{sec:design}

\begin{figure*}[t!]
\centering
\includegraphics[width=0.95\linewidth]{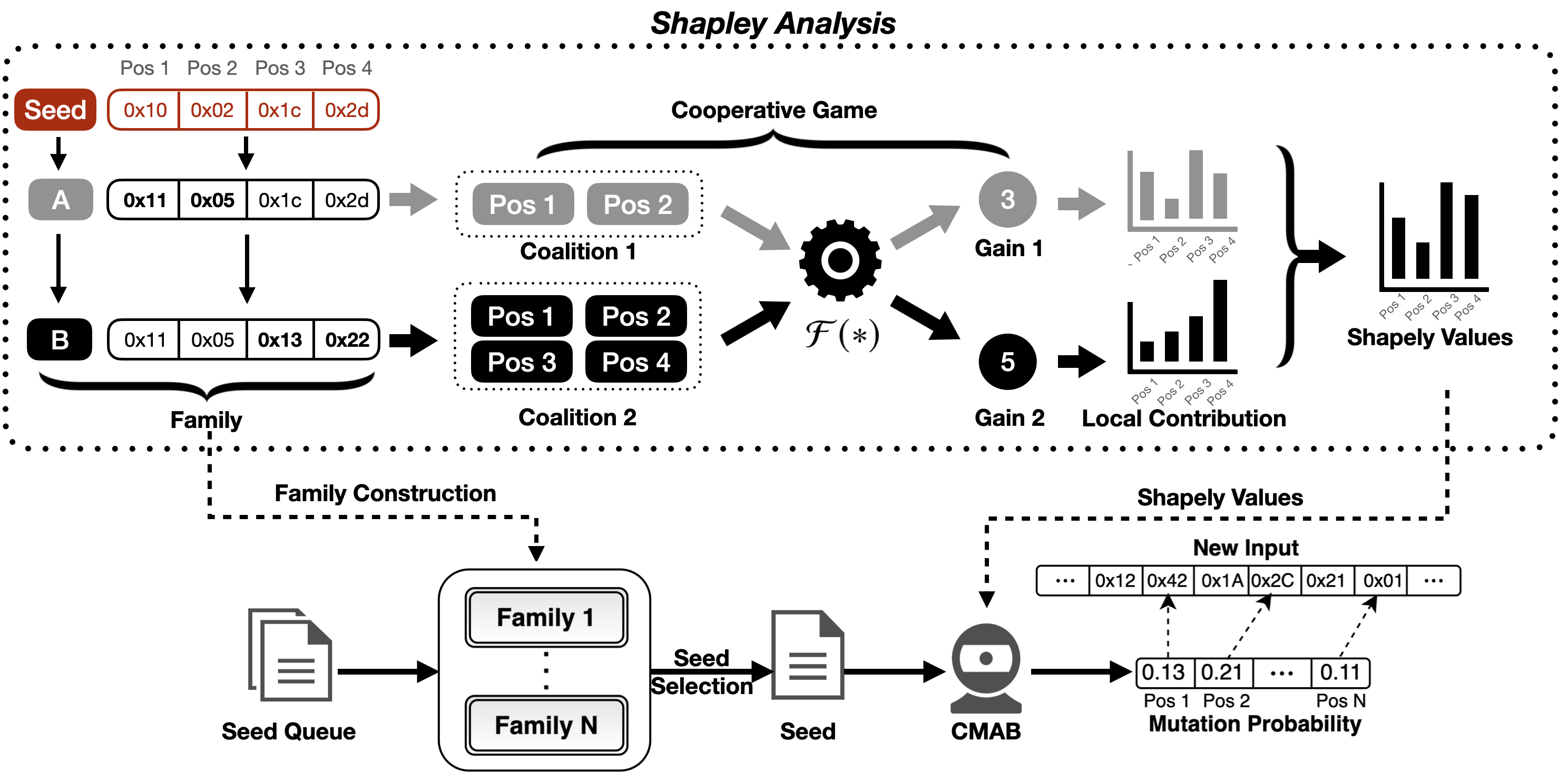}
\caption{The Framework of \tool{}. It groups seeds into families, calculates Shapely values, and utilizes CMAB to optimize Shapley-guided byte mutation.
}
\label{Framework}
\vspace{-2mm}
\end{figure*}

Our \tool{} focuses on the byte mutation and quantifies the importance of byte positions.
We understand the byte mutation across multiple seeds as a process of Shapley analysis and use the Shapley value to quantify the contribution of each byte.
Figure~\ref{Framework} shows the framework of our \tool{}, which uses Shapley values of bytes to boost fuzzing efficiency.
We treat the schedule of bytes on the seed as the cooperative game and calculate the Shapley value for each byte.
To relieve the overhead of Shapley analysis, we deduce that the seeds, which are retained from the same original seed and do not change the length, belong to the same cooperative game and share the same Shapley values for bytes.
Thus, we group such seeds into a family to calculate the shared Shapley values in each family.

Fuzzing efficiency can be improved by assigning more energy to the bytes with higher Shapley values. 
However, we need to mutate all bytes and generate all possible inputs to get an accurate Shapley Value during fuzzing.
To make the trade-off between mutating high Shapley value bytes and low-frequently chosen bytes, \tool{} uses CMAB to build a probability distribution for optimizing this trade-off.

\subsection{Shapley Analysis Among Bytes} \label{subsec:cg_all}

As mentioned, bytes in a seed are collaboratively mutated to discover new edges. Therefore, we formulate the process of byte mutation as a cooperative game and use Shapley value to quantify the contribution of each byte. We first formulate the mutation of one seed and then extend it to multiple seeds based on the characteristics of mutation.


\subsubsection{Shapley Value} \label{subsec:shapval}

The Shapley value is a solution used in Cooperative Game Theory that fairly distributes gains among several players working in a coalition. 
A cooperative game is one that $N$ players form coalitions and obtain gains by a characteristic function.
Shapley Value can be used to quantify the contribution of each player in a game.
In the Cooperative Game Theory,
the Shapley Value $\phi_{j}$ of the player $x_{j}$ to the coalition can be calculated as~\cite{Shapley}:
\begin{equation} \label{equ:shapely}
    \phi_{j}=\sum_{G \subseteq X \backslash \{x_{j}\}} \frac{|G| !(N-|G|-1) !}{N !}(\mathcal{F}(G \cup \{x_{j}\})-\mathcal{F}(G)),
\end{equation}
where there are $N$ players in the game, and a coalition $G$ is formed by $m (m \leq N)$ players.
$X$ is the set of all players and $X \backslash \{x_{j}\}$ is all possible coalitions of players excluding the player $x_{j}$. $|G|$ is the number of players in the set $G$.
$\mathcal{F}(\ast)$ is a characteristic function that describes the gain $R$ that the coalition of players can obtain, \ie $R = \mathcal{F}(G)$.
$\mathcal{F}(G \cup \{x_{j}\})-\mathcal{F}(G)$ is the marginal contribution of the player $x_{j}$ on the coalition $G$, which describes the effect on the gain by the player $x_{j}$. In other words, the Shapley value $\phi_{j}$ is a weighted average of $x_{j}$'s marginal contributions across all possible player coalitions. 
In other words, the Shapley value reflects the importance of the player $x_j$ in the game.

Coverage-guided fuzzing usually has some initial seeds, and mutation of them may generate new seeds.
If an input generated by mutating a seed discovers new edges comparing to the overall discovered coverage, the input is retained as a new seed.
For the following fuzzing trials, the new seed will be selected and mutated to generate new inputs. 
This process of generating new inputs and testing them can be regarded as the process of Shapley analysis, \ie \textit{trying possible combinations of bytes}.
Yet, the problem is how to gather the trials in fuzzing so that these trials belong to the same Shapley analysis.


\subsubsection{Shapley Analysis in One Seed} \label{subsec:cg_oneseed}

For a seed, the \textbf{cooperative game} is the schedule of bytes on the seed, including the selection of some certain bytes and the mutation of the selected bytes. This game finds a total of $n$ edges and the Shapley value is utilized to analyze the contribution of each byte pertaining to the discovery of those $n$ edges. Thus, a \textbf{player} in the game is a byte in the seed $S_0$, and the \textbf{coalition} is that some certain bytes are mutated together. Since coverage-guided fuzzing aims to discover new code, we define the \textbf{gain} as the number of \textit{self-new edges} discovered by an input $i$ generated by mutating the seed $S_0$. 
To ensure that the gain is consistent during fuzzing, the \textit{self-new edges} are defined as the new edges when comparing the edges discovered by the input $i$ and the initial seed $S_0$. From the perspective of program logic, \textit{a byte related to more path constraints will have larger contribution to the code discovery}. Thus, the use of self-new edges can provide more information about the relations between input bytes and path constraints.
For most fuzzers, such as AFL-based fuzzers~\cite{aflchurn, Zhu2020CSI}, new edges are obtained by comparing to the overall coverage. However, this kind of new edges is ``stateful'' because the order of mutation can influence the result. As a result, depending on the time of mutation, the gain of the same mutation may be different. Therefore, we use self-new edges to avoid the variation of gain. Finally, the \textbf{characteristic function} $\mathcal{F}(*)$ is the mapping between the collaborative mutation and the number of self-new edges.
With the formulation of byte mutation, we can use Equation~(\ref{equ:shapely}) to calculate the Shapley value for each byte. Since the Shapley value reflects the importance of a byte pertaining to code discovery, we can improve the efficiency of fuzzing by \textit{assigning more energy to the bytes with higher Shapley values}.

\subsubsection{Shapley Analysis Across Seeds} \label{subsec:cg_multiseed}
Multiple seeds are possible to be treated as one seed from the perspective of Shapley analysis.
Although Section~\ref{subsec:cg_oneseed} formulates byte mutation based on one seed $S_0$, the process of performing Shapley analysis will generate new seeds. 
If the mutation does not change the length of the original seed $S_0$, the new seed $S_1$ is actually one of the combinations in the Shapley analysis of seed $S_0$. 
If the length changes, the relative positions of bytes will change,
as well as the number of players in the Shapley analysis.
Therefore, without changing length, any mutation of seed $S_1$ still belongs to the Shapley analysis of seed $S_0$. 
For example, if seed $S_0$ mutates byte $i$ to get seed $S_1$, and $S_1$ mutates byte $j$ to get seed $S_2$.
The lengths of the seeds $S_0$, $S_1$, and $S_2$ are the same.
Then, the seed $S_2$ can be obtained by mutating bytes $i,j$ in seed $S_0$.
Therefore, the seeds, which are retained from the same original seed and do not change the length, are part of the combinations for the original seed.
We group such seeds (\eg seeds $S_0$, $S_1$ and $S_2$) into a \textit{family} so that seeds in the same family participate in the same cooperative game. 
To conclude, seeds in a family 1) inherit from the same original seed, 2) obtain the subset of players in Equation~(\ref{equ:shapely}) by comparing to the original seed, and 3) have the same length.

The family can reduce the overhead of Shapley analysis because it gathers different seeds into the same Shapley analysis.
During fuzzing, all seeds will be selected and mutated. If we treat byte mutation in every seed as an individual Shapley analysis, it will have too much overhead.
Because seeds in the same family belong to the same Shapley analysis, mutations of all seeds in the family are used to calculate the Shapley values for the first seed in the family. 
The Shapley values for bytes of the first seed are shared with all other seeds in the same family.

\subsection{Temporary Shapley Value Update} \label{sec:shap_update}
In this section, we transform the calculation of Shapley value into a form that can be updated during fuzzing. We then reduce the overhead of calculating Shapley values based on the properties of fuzzing. 

\subsubsection{Transformation of Shapley Value} \label{subsec:trans_shap}
Since a byte includes both position (the order in an input) and value, all possible combinations of byte mutation include combinations of byte positions coupled with combinations of byte values.
In Equation~(\ref{equ:shapely}), when the subset $G$ of bytes is selected, the byte positions are fixed but the byte values have different combinations. Therefore, the gain of the subset $G$ comes from all the combinations of byte values. In fuzzing, a combination of byte values is an input generated by mutating the selected positions of $G$. Moreover, the gain of subset $G$ is obtained by accumulating gains of all inputs belonging to $G$. Then, we have
\begin{equation} \label{equ:sum_value}
    \mathcal{F}(V) = \sum_{v\in V} \mathcal{F}(\{v\}),
\end{equation}
where $V$ is the set of all possible combinations of byte values in $G$, $v$ (an input) is a combination belonging to $V$, and $\{v\}$ is a set that only have one element $v$. 
With equations~(\ref{equ:shapely}) and (\ref{equ:sum_value}), let $k(G)=\frac{|G| !(N-|G|-1) !}{N !}$ and $P$ be the positions selected in $G$, then we have
\begin{equation} \label{equ:shapely_sum}
\begin{split}
\phi_{j} & = \sum_{G \subseteq X \backslash \{x_{j}\}} k(G) \bigl( \mathcal{F}(G \cup \{x_{j}\})-\mathcal{F}(G) \bigr) \\
         & = \sum_{P \subseteq X \backslash \{x_{j}\}} k(P) \Bigl( \sum_{v\in V} \mathcal{F}(\{v\}\cup\{x_j\}) - \sum_{v\in V} \mathcal{F}(\{v\}) \Bigr) \\
         & = \sum_{P \subseteq X \backslash \{x_{j}\}} k(P) \biggl( \sum_{v\in V} \Bigl( \mathcal{F}(\{v\}\cup\{x_j\}) - \mathcal{F}(\{v\}) \Bigr) \biggr) ,\\
         & = \sum_{P \subseteq X \backslash \{x_{j}\}} \sum_{v\in V} k(P)\Bigl( \mathcal{F}(\{v\}\cup\{x_j\}) - \mathcal{F}(\{v\}) \Bigr) \\
         & = \sum_{P \subseteq X \backslash \{x_{j}\}} \sum_{v\in V} \varphi_{j,v}
\end{split}
\end{equation}
where 
\begin{equation}\label{equ:shap_eta}
    \varphi_{j,v} = k(P)\Bigl( \mathcal{F}(\{v\}\cup\{x_j\}) - \mathcal{F}(\{v\}) \Bigr)
\end{equation}

Equation~(\ref{equ:shapely_sum}) indicates that the Shapley value of the byte $x_j$ can be accumulated by analyzing individual inputs. Specifically, each time an input $v$ discovers self-new edges, we can use $\varphi_{j,v}$ to calculate the local contribution of byte $x_j$ based on the input. Then, when more inputs are generated, we accumulate their $\varphi$ to update the Shapley value of byte $x_j$. To simplify the computation, we set $k(P)$ to 1.
Therefore, we can use Shapley value to guide fuzzing without changing the most fuzzing logic described in Section~\ref{subsec:shapval}. 
Notably, the contribution of a byte $x_j$ has the same meaning to the contribution of its position $j$ because positions are fixed when bytes have been selected.

\begin{figure}[t!]
\centering
\includegraphics[width=0.8\linewidth]{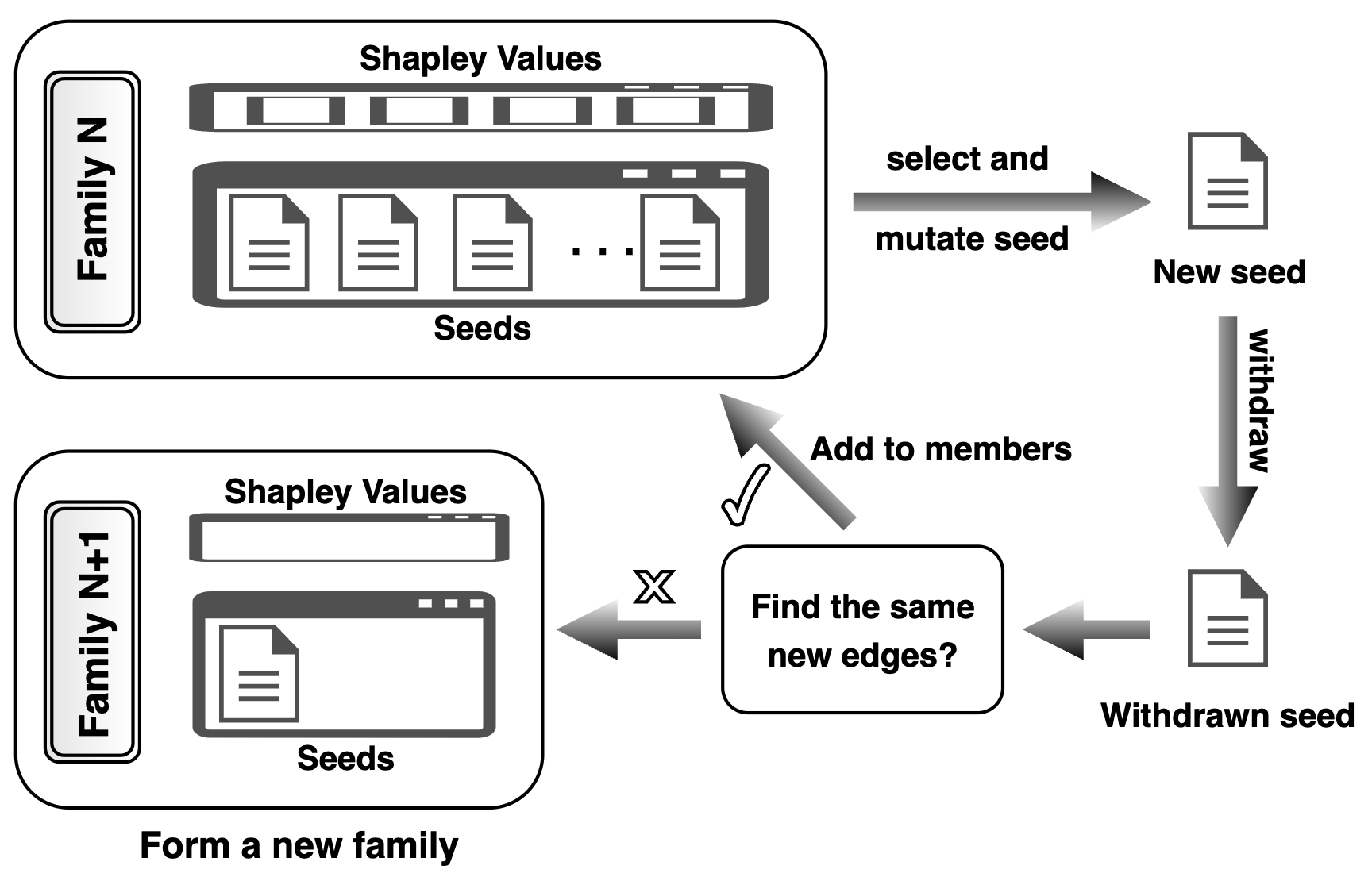}
\caption{The update of family and family members.}
\label{fig:family_update}
\end{figure}

\subsubsection{Shapley Update During Fuzzing}
As mentioned in Section~\ref{subsec:cg_multiseed}, Shapley values are calculated for the original seed in a family, which can significantly improve the efficiency of fuzzing. To further improve the efficiency of fuzzing, we update Shapley values by 1) ignoring the generated inputs that do not discover self-new edges, and 2) reducing the combinations of bytes when self-new edges are discovered.
If a mutation does not discover self-new edges, then $\mathcal{F}(\ast) = 0$. Therefore, the Shapley value remains the same and we do not need to update it.
If a cooperative mutation discovers self-new edges, we remove the redundant bytes that do not influence the result of self-new edges. If fuzzing mutates bytes $M_{b} = \{b_1, b_2, ..., b_n\}$ to discover self-new edges, for a redundant byte $b_i$, we get $\mathcal{F}(M_b\backslash \{b_i\}) = \mathcal{F}(M_{b})$. This immediately derives that the value of the related $\varphi$ is zero. Therefore, we do not need to update Shapley values for redundant bytes.
For the necessary bytes that influence the result of self-new edges, denoted as $N_b$, exclusion of any byte $b_j$ in $N_b$ will not discover self-new edges, \ie $\mathcal{F}(N_b\backslash \{b_j\}) = 0$. 
Therefore, all necessary bytes in this mutation have the same value of $\varphi$, indicating that we can only calculate $\varphi$ for one necessary byte to update Shapley values for all necessary bytes. 
In fuzzing, when self-new edges are discovered, we recover mutations of bytes to check whether those bytes are redundant.
If the recovery of the mutation of a byte does not change the result of self-new edges, the byte is a redundant byte. To simplify the computation, we consider the mutated bytes, excluding the redundant bytes, as necessary bytes.

\subsubsection{Family Update}
The family for seeds can reduce the overhead of calculating Shapley values because Shapley values for one seed can be shared with multiple seeds.  
The key of building a family is that the mutation does not change the length of original seeds. Yet, some mutators, such as \textit{deletion of bytes}, in fuzzing will change the length of seeds. 
To enrich family members, \tool{} will withdraw all mutators that change length when new code are discovered.
If the result can still find the same new edges after withdrawing those mutators, \tool{} will retain the result of withdrawing to the family.
For example, as shown in Figure~\ref{fig:family_update}, we select a seed from family $N$ and mutate it. 
If an input $i_1$ discovers new edges, \tool{} will withdraw all mutators that change length, which generates another input $i_2$. If $i_2$ can still find the same new edges that $i_1$ finds, \tool{} retains $i_2$ as a new seed and adds it to the family $N$.
Otherwise, $i_1$ will form a new family $N+1$ if mutation of $i_1$ can further discover new edges.
As a result, \tool{} has a reasonable number of families and a number of family members.

\subsection{Shapley-guided Byte Selection}
\label{subsec:mutate}
The insight from Section~\ref{sec:motivation} indicates that more energy can be assigned to bytes with higher Shapley values. 
In this section, we present Shapley-guided byte selection while the mutation of byte values is performed by AFL++.
However, since the Shapley value is calculated based on part of all possible inputs during fuzzing, it cannot guarantee that the bytes with high Shapley values can always maintain high.
Besides, seeds share the byte importance with the original seed in the same family.   
Although a byte may have a high Shapley value, the difficulty of solving path constraints varies in different execution paths.
When some bytes are frequently selected but with no new edges discovered, we do not want to waste more energy on those bytes.
Therefore, we need to balance the trade-off between mutating high Shapley value bytes and low-frequently chosen bytes.

To maximize code discovery, \tool{} uses Contextual Multi-armed Bandit (CMAB)~\cite{CMAB, linucb} to achieve the trade-off in using Shapley values.
Based on the experience of playing arms, CMAB tries to maximize the rewards of playing multi-armed bandits, taking into account the personality of a player (\ie the context). The trade-off in CMAB is between 1) focusing on the arms with higher historical performance and 2) updating rewards for all arms. This exactly fits into our problem of selecting bytes based on Shapley values.

With Shapley values, \tool{} uses CMAB to build a probability distribution for selecting bytes.
To build the probability distribution, CMAB calculates scores for each byte. The bytes with higher scores are assigned with higher probabilities. Given a seed $s$, the feature vector $f_{s}$ extracted from the corresponding execution path (\ie the context), and a byte $p$, the score of the byte is calculated as:
\begin{equation} \label{equ:score_cal}
Score(s,p) = E[\phi_{p}|f_{s}] + 0.5 U_{s,p},
\end{equation}
where $E[\phi_{p}|f_{s}]$ is the expected reward, $U_{s,p}$ is the upper confidence bound, and $\phi_{p}$ is the Shapley value of byte $p$. As mentioned, the same $\phi_p$ in different execution paths has different influence on discovering new code. Therefore, we use $f_s$ to represent the context of $\phi_{p}$, and $f_s$ is extracted based on its edges in the execution path.
First, we find ten center seeds in the seed corpus whose paths are as far as possible from each other.
Then, for a seed $s$, we compute the cosine similarity between the path exercised by that seed and the paths exercised by ten center seeds to obtain a fixed dimensional vector $f_s$.

To calculate $E[\phi_{p}|f_{s}]$ and $U_{s,p}$, the byte $p$ needs to maintain a tuple $(D_{p}, R_{p})$, where $D_p$ is a matrix about all history feature vectors and $R_p$ is a vector about Shapley values. 
To update the tuple $(D_{p}, R_{p})$:
\begin{equation}
\begin{split}
    D_p & \gets D_p + f_sf^T_s \\
    R_p & \gets R_p + \phi_p f_s
\end{split}
\end{equation}

\textit{Expected Reward.} In our model, the expected reward can be calculated by:
\begin{equation} \label{equ:E_calculation}
E[\phi_{p}|f_{s}] = f_{s}^{T}\hat{\theta},
\end{equation}
where $\hat{\theta}$ is the best coefficient vector, and it can be estimated by applying ridge regression to the tuple $(D_{p}, R_{p})$:
\begin{equation} 
\hat{\theta} = (D_{p}^{T}D_{p} + I_{k})^{-1} D_{p}^{T} R_{p},
\end{equation}
where $I_{k}$ is a $k$-dimensional identity matrix, and $k$ is the dimension of the feature vector $f_s$.

\textit{Upper Confidence Bound.}
The second part of the score is the upper confidence bound $U_{s,p}$, which can be calculated by~\cite{walsh2009exploring}:
\begin{equation} \label{equ:U_calculation}
U_{s,p} = \sqrt{f_{s}^{T} (D_{p}^{T}D_{p} + I_{d})^{-1} f_{s}}
\end{equation}
The more times a byte is selected, the smaller its $U_{s,p}$ is. Thus, the $U_{s,p}$ intends to assign more energy to the bytes that have been less frequently selected, \ie the bytes with low Shapley values.
When bytes with high Shapley values are selected frequently, their scores will be decreased, as well the selection probability.
As a result, the score calculated by Equation~(\ref{equ:score_cal}) achieves the trade-off for Shapley-guided fuzzing.

The scores calculated for bytes are further utilized to generate probabilities for selecting bytes. A byte with a higher score is assigned a higher probability. For each mutation, \tool{} first randomly sets a number $n$, which is the number of times to perform sub-mutations. That is, it will perform the sub-mutation by $n$ times in one mutation. Then, each time the fuzzer decides which byte to be sub-mutated, the fuzzer will use the probability distribution to select.



\subsection{Threats to Validity}
While our approach has shown promising performance and effectiveness, there are still certain potential factors that could constrain the practical application of \tool{}.

\subsubsection{Overhead of Shapley Calculation}
\tool{} models the process of byte mutation as a cooperative game and uses Shapley analysis to quantify the contribution of each byte.
Assume that self-new edges are found in a mutation, and the number of its mutated positions is $N$.
To calculate the Shapley value, we need to analyze the number of self-new edges found in $2^{N}$ subsets of mutated positions.
To reduce the overhead, \tool{} uses two special cases to simplify the process of Shapley analysis.
Firstly, if there are redundant bytes that do not influence the result of the self-new edge, we remove these bytes to reduce $N$.
Secondly, if the mutated bytes are necessary for the discovery of self-new edges, these bytes have the same Shapley value.
However, if these two conditions are not met and $N$ is large, some overhead will be introduced.
In the future, we can use lightweight and improved algorithms of Shapley analysis to alleviate this problem.

\subsubsection{Changeable Seed Length}
Moreover, \tool{} builds a family system to alleviate the overhead brought by Shapley analysis. Within a family, all seeds in the same family share the same Shapley values for bytes.
However, when maintaining a family, we ignore mutated inputs that change length.
This is because the use of mutators that change seed length may disorder byte positions in seeds, and we cannot achieve position mapping between seeds.
We can study the mathematical theory for the Shapley analysis of the changeable length in the future.
In the existing version, if a new seed is generated by changing length, we use this seed to create a new family to alleviate this problem.

\subsubsection{The Impact of Self-New Edges}
As new edge discovery becomes progressively more difficult over time, \tool{} utilizes self-new edges as a metric to measure gains and identify bytes that may impact multiple branches. Nevertheless, in some cases, bytes that contribute to the self-new edges may not necessarily lead to the discovery of extra new edges.
The self-new edges might have already been discovered by other seeds.
Therefore, bytes with high Shapley values could actually contribute less than what the Shapley values indicate. If we allocate too much energy to these bytes, it could lead to a decrease in fuzzing efficiency. For example, suppose byte $i$ in a seed is related to $y$ edges. If all these $y$ edges have been explored, we can deduce that byte $i$ has made a significant contribution. However, since all the edges related to byte $i$ have been explored, allocating any amount of energy to byte $i$ will not help discover new edges.


\subsubsection{Semantic Consistency Within Families}
To reduce the overhead of calculating Shapley values, the seeds in a family share the same Shapley values. However, despite implementing a strict family construction method (seeds with genetic relationships and the same lengths), there may still be cases where the semantics between seeds are inconsistent.
Semantic inconsistency can lead to bytes with different meanings sharing the same Shapley value, thereby incorrectly estimating the importance of bytes. For instance, a length field can often determine the semantics of the following bytes. In such a case, our method cannot guarantee semantic consistency between seeds. Suppose there is a seed $A$, where a byte $i$ is a length field. Seed $B$ is derived from the seed $A$ by mutating the byte $i$. As the change in the length field directly alters the meaning of the subsequent bytes, it is likely that a byte $j$, which follows the byte $i$, in the seed $A$ and seed $B$ (with the same position) carries different meanings. This means that even if seeds $A$ and $B$ have the same length and genetic relationship, the byte $j$ in seed $A$ and seed $B$ is not semantically consistent. If we continue to share the Shapley value of the byte $j$ among these two seeds, it would result in inaccurate Shapley values, enabling some inefficient bytes to be allocated with higher mutation energy, thereby affecting fuzzing efficiency.

\section{Evaluation} \label{sec:eval}

We implement our \tool{} based on AFL++~\cite{aflpp}, which integrates many state-of-the-art fuzzing algorithms. We mainly modify the mutation process and the byte schedule. 
In the evaluation, we evaluate our \tool{} with byte-scheduling fuzzers and commonly used fuzzers.
To ensure fair results, we conduct our experiments on two third-party testing platforms, UNIFUZZ~\cite{unifuzz} and MAGMA~\cite{magma}. Our experiments are performed on a system running Ubuntu 18.04 with 103 cores (Intel(R) Xeon(R) Gold 6230R CPU) and 256 GB of memory. Each fuzzer uses one CPU core to run one target program.

\begin{table}[!t]
\centering
\caption{Target programs from UNIFUZZ.}\label{programsUNIFUZZ}
\resizebox{0.9\linewidth}{!}{
\begin{threeparttable}
\begin{tabular}{llll}
    \toprule
    \textbf{Targets} & \textbf{Projects} & \textbf{Input formats} & \textbf{Test instructions} \\ \midrule
    \textbf{tiff2bw\tnote{2}} & \multirow{3}{*}{libtiff 3.9.7} & \multirow{3}{*}{tiff} & tiff2bw @@\tnote{1}~{ } /dev/null \\
    \textbf{tiffsplit} &&& tiffsplit @@ /dev/null\\
    \textbf{tiff2pdf\tnote{2}} &&& tiff2pdf @@ /dev/null\\
    \midrule
    \multirow{3}{*}{\textbf{nm}} & \multirow{6}{*}{binutils-2.28} & \multirow{4}{*}{ELF} & (-A -a -l -S -s \\
    &&&\ \ \ --special-syms --synthetic\\
    &&&\ \ \ --with-symbol-versions -D @@) \\
    \textbf{objdump} &&& -S @@ \\
    \textbf{readelf\tnote{2}} &&& -a @@ \\
    \textbf{size\tnote{2}} &&& @@ \\
    \midrule
    \textbf{pdftotext} & pdftotext 4.00 & pdf & @@ /dev/null \\
    \midrule
    \textbf{infotocap} & jasper 2.0.12 & text & -o /dev/null @@ \\
    \midrule
    \textbf{mp42aac} & Bento4 1.5.1-628 & mp4 & @@ /dev/null \\
    \midrule
    \multirow{2}{*}{\textbf{tcpdump}}
     & tcpdump 4.8.1 + & \multirow{2}{*}{tcpdump100} & \multirow{2}{*}{-e -vv -nr @@} \\
     & libpcap 1.8.1 && \\
    \midrule
    \textbf{flvmeta} & flvmeta 1.2.1 & flv &  @@ \\
    \midrule
    \textbf{lame3.99.5} & lame 3.99.5 & wav & @@ /dev/null \\
    \midrule
    \textbf{imginfo} & ncurses 6.1 & jpg & -f @@ \\
    \midrule
    \textbf{mujs} & mujs 1.0.2 & js & @@ \\
    \midrule
    \textbf{exiv2} & exiv2 0.26 & jpg & @@ \\
    \bottomrule
\end{tabular}
\begin{tablenotes}
\item[1] @@: A placeholder indicating that the input is a file.
\item[2] \textbf{tiff2bw, tiff2pdf, readelf} and \textbf{size} are not from UNIFUZZ
\end{tablenotes}
\end{threeparttable}

}
\end{table}


\noindent\textbf{Evaluation Metrics.}
We use \textit{afl-showmap} in AFL++ to count the number of edges and compile target programs with ASAN~\cite{asan} enabled. When analyzing unique bugs, we use the top-three rule to de-duplicate bugs~\cite{Klees2018Evaluating}.

\noindent\textbf{Target Programs and Initial Seeds.}
We select 16 commonly used programs for testing from UNIFUZZ and other papers, as shown in the Table \ref{programsUNIFUZZ}.
For each target program, all fuzzers use the same initial seeds, which are collected from UNIFUZZ, FuzzBench\cite{fuzzbench} and public seed corpus. 
Most target programs have multiple initial seeds as suggested by Adrian Herrera~\cite{herrera2021seed}.

\begin{table*}[!t]
\centering
\caption{Edge Coverage and Analysis Time comparison between different fuzzers (on \texttt{ALL SEEDS}).}
\resizebox{\linewidth}{!}{
\begin{threeparttable}
\begin{tabular}{@{\extracolsep{-2pt}}cc|ccc|ccc|ccc|ccc|ccc|ccc@{}} 
\toprule
\multicolumn{2}{c}{\textbf{Programs}} & \multicolumn{3}{c}{\textbf{\tool{}}} & \multicolumn{3}{c}{\textbf{GreyOne}} & \multicolumn{3}{c}{\textbf{ProFuzzer}} & \multicolumn{3}{c}{\textbf{Angora}} & \multicolumn{3}{c}{\textbf{PreFuzz}} & \multicolumn{3}{c}{\textbf{NEUZZ}}\\
\textbf{Name} & \textbf{Len}& \textbf{Cov.} & \textbf{Time} & \textbf{\#Bug}& \textbf{Cov.} & \textbf{Time} & \textbf{\#Bug}& \textbf{Cov.} & \textbf{Time} & \textbf{\#Bug}& \textbf{Cov.} & \textbf{Time} & \textbf{\#Bug}& \textbf{Cov.} & \textbf{Time} & \textbf{\#Bug}& \textbf{Cov.} & \textbf{Time} & \textbf{\#Bug}\\
\midrule

tiff2pdf & 448 & 4401 & 85s & 0 & 4486 & 35207s & 0 & 4578 & 12694s & 1 & 2314 & 235s & 0 & 3475 & G & 1 & 2832 & G\tnote{1}~ & 1 \\ 
lame & 13818 & 3656 & 3985s & 5 & 3645 & 80225s & 4 & 3649 & 70003s & 4 & 2265 & 5813s & 4 & - & - & 0 & -\tnote{2} & - & 0 \\ 
readelf & 272030 & 5611 & 1048s & 4 & 5058 & 78788s & 3 & 5168 & 74878s & 6 & 5786 & 3758s & 6 & - & - & 0 & - & - & 0 \\ 
exiv2 & 25633 & 3790 & 135s & 15 & 3626 & 35401s & 5 & 3698 & 32311s & 12 & 4291 & 615s & 11 & 2866 & G & 0 & - & - & 0 \\ 
flvmeta & 16454 & 230 & 0s & 2 & 230 & 32782s & 2 & 230 & 11625s & 2 & 230 & 201s & 2 & - & - & 0 & - & - & 0 \\ 
nm & 272030 & 3136 & 248s & 17 & 2628 & 83058s & 9 & 2750 & 81747s & 10 & 2657 & 2351s & 51 & - & - & 0 & - & - & 0 \\ 
tiffsplit & 10032 & 1709 & 5s & 7 & 1699 & 13398s & 5 & 1719 & 6220s & 6 & 890 & 291s & 1 & 1189 & G & 2 & 1113 & G & 1 \\ 
tiff2bw & 10032 & 1839 & 87s & 6 & 1870 & 12709s & 6 & 1848 & 8702s & 5 & 1180 & 310s & 1 & 1651 & G & 0 & 1441 & G & 0 \\ 
objdump & 272030 & 4958 & 742s & 14 & 3864 & 79211s & 4 & 4106 & 83913s & 5 & 3192 & 3672s & 3 & - & - & 0 & - & - & 0 \\ 
pdftotext & 12465 & 6613 & 3508s & 23 & 5794 & 80548s & 2 & 5777 & 80591s & 3 & 4250 & 14801s & 2 & 4902 & G & 5 & 4394 & G & 0 \\ 
mp42aac & 31988 & 1266 & 19s & 2 & 1144 & 75949s & 0 & 1166 & 66141s & 0 & 1120 & 540s & 0 & 1017 & G & 0 & 950 & G & 0 \\ 
tcpdump & 6983 & 12764 & 328s & 1 & 11558 & 50599s & 1 & 11879 & 29383s & 0 & 7615 & 2510s & 3 & 4672 & G & 0 & 6019 & G & 0 \\ 
mujs & 6983 & 4136 & 0s & 0 & 4020 & 18157s & 0 & 3979 & 7181s & 0 & 2364 & 8829s & 0 & 2432 & G & 0 & 2393 & G & 0 \\ 
size & 272030 & 1860 & 188s & 0 & 1667 & 77562s & 0 & 1690 & 78671s & 0 & 1988 & 2032s & 0 & - & - & 0 & - & - & 0 \\ 
infotocap & 2519 & 1817 & 276s & 7 & 1670 & 39193s & 6 & 1530 & 44335s & 5 & 940 & 1104s & 0 & 1266 & G & 1 & 1071 & G & 0 \\ 
imginfo & 2519 & 1895 & 30s & 0 & 1818 & 36298s & 0 & 1744 & 43274s & 0 & 1384 & 184s & 0 & - & - & 0 & - & - & 0 \\
\midrule
\textbf{Total} & ~  & \textbf{59681} & & \textbf{103} & 54777 & & 47 & 55511 & & 59 & 42466 & & 84 & 23470 & & 9 & 20213 & & 2 \\

\bottomrule
\end{tabular}
\begin{tablenotes}[para]
\small
\item[1] G: A GPU is required to train the model.
\item[2]-: The fuzzer fails to run on this program due to large input size.
\end{tablenotes}
\end{threeparttable}
}

\label{all_seeds}
\end{table*}

\begin{table*}[!t]
\centering
\caption{Edge Coverage and Analysis Time comparison between different fuzzers (on \texttt{SEEDS<10000}).}
\resizebox{\linewidth}{!}{
\begin{threeparttable}
\begin{tabular}{@{\extracolsep{-2pt}}cc|ccc|ccc|ccc|ccc|ccc|ccc@{}} 
\toprule
\multicolumn{2}{c}{\textbf{Programs}} & \multicolumn{3}{c}{\textbf{\tool{}}} & \multicolumn{3}{c}{\textbf{GreyOne}} & \multicolumn{3}{c}{\textbf{ProFuzzer}} & \multicolumn{3}{c}{\textbf{Angora}} & \multicolumn{3}{c}{\textbf{PreFuzz}} & \multicolumn{3}{c}{\textbf{NEUZZ}}\\
\textbf{Name} & \textbf{Len}& \textbf{Cov.} & \textbf{Time} & \textbf{\#Bug}& \textbf{Cov.} & \textbf{Time} & \textbf{\#Bug}& \textbf{Cov.} & \textbf{Time} & \textbf{\#Bug}& \textbf{Cov.} & \textbf{Time} & \textbf{\#Bug}& \textbf{Cov.} & \textbf{Time} & \textbf{\#Bug}& \textbf{Cov.} & \textbf{Time} & \textbf{\#Bug}\\
\midrule

tiff2pdf & 448 & 4461& 30s & 1 & 4450& 39189s & 0 & 4552& 15278s & 0 & 2306& 50s & 0 & 3466& G & 0 & 2543& G\tnote{1}~ & 0 \\ 
readelf & 4230 & 5359& 77s & 5 & 5346& 39952s & 6 & 5272& 38537s & 3 & 5554& 1628s & 7 & 4540& G & 6 & 4208& G & 4 \\ 
exiv2 & 8437 & 3882& 196s & 7 & 3385& 50513s & 4 & 3691& 31915s & 3 & 3903& 2365s & 9 & 3881& G & 0 & 3151& G & 0 \\ 
nm & 4230 & 2822& 17s & 30 & 2818& 29069s & 20 & 2884& 13905s & 22 & 2409& 2056s & 35 & 2260& G & 14 & 1749& G & 0 \\ 
tiffsplit & 7222 & 1718& 4s & 6 & 1725& 13541s & 6 & 1715& 4519s & 7 & 919& 219s & 3 & 1203& G & 2 & 1128& G & 0 \\ 
tiff2bw & 7222 & 1882& 42s & 7 & 1876& 12031s & 6 & 1808& 6040s & 6 & 1186& 200s & 1 & 1594& G & 0 & 1489& G & 0 \\ 
objdump & 4230 & 4654& 104s & 16 & 4145& 75874s & 10 & 4414& 50760s & 13 & 3085& 4182s & 13 & 3560& G & 4 & 3136& G & 0 \\ 
tcpdump & 2305 & 12963& 227s & 1 & 12164& 40025s & 1 & 12497& 13022s & 0 & 7826& 2119s & 1 & 7963& G & 1 & 6251& G & 0 \\ 
mujs & 2305 & 4145& 0s & 0 & 4000& 19877s & 0 & 4015& 7078s & 0 & 2383& 8677s & 0 & 2589& G & 1 & 2388& G & 0 \\ 
size & 4230 & 1795& 22s & 0 & 1785& 12623s & 0 & 1812& 9547s & 0 & 1794& 1259s & 1 & 1458& G & 0 & 1224& G & 0 \\ 
infotocap & 2519 & 1807& 457s & 8 & 1767& 38278s & 5 & 1399& 45154s & 2 & 840& 1647s & 0 & 1159& G & 1 & 1074& G & 0 \\ 
imginfo & 2519 & 2486& 9s & 0 & 1744& 26514s & 0 & 1680& 24268s & 0 & 1413& 244s & 0 & 1464& G & 0 & 1176& G & 0 \\
\midrule
\textbf{Total} &  & \textbf{47974} & & \textbf{81} & 45205 & & 58 & 45739 & & 56 & 33618 & & 70 & 35137 & & 29 & 29517 & & 4 \\

\bottomrule
\end{tabular}
\begin{tablenotes}[para]
\small
\item[1]G: A GPU is required to train the model. 
\end{tablenotes}
\end{threeparttable}
}
\label{seeds_10000}
\end{table*}

\begin{table*}[!t]
\centering
\caption{Edge Coverage and Analysis Time comparison between different fuzzers (on \texttt{SEEDS<1000}).}
\resizebox{\linewidth}{!}{
\begin{threeparttable}
\begin{tabular}{@{\extracolsep{-2pt}}cc|ccc|ccc|ccc|ccc|ccc|ccc@{}} 
\toprule
\multicolumn{2}{c}{\textbf{Programs}} & \multicolumn{3}{c}{\textbf{\tool{}}} & \multicolumn{3}{c}{\textbf{GreyOne}} & \multicolumn{3}{c}{\textbf{ProFuzzer}} & \multicolumn{3}{c}{\textbf{Angora}} & \multicolumn{3}{c}{\textbf{PreFuzz}} & \multicolumn{3}{c}{\textbf{NEUZZ}}\\
\textbf{Name} & \textbf{Len}& \textbf{Cov.} & \textbf{Time} & \textbf{\#Bug}& \textbf{Cov.} & \textbf{Time} & \textbf{\#Bug}& \textbf{Cov.} & \textbf{Time} & \textbf{\#Bug}& \textbf{Cov.} & \textbf{Time} & \textbf{\#Bug}& \textbf{Cov.} & \textbf{Time} & \textbf{\#Bug}& \textbf{Cov.} & \textbf{Time} & \textbf{\#Bug}\\
\midrule

tiff2pdf & 448 & 4383 & 99s & 1 & 4493 & 39407s & 0 & 4449 & 12856s & 1 & 2459 & 58s & 0 & 3398 & G & 1 & 2709 & G\tnote{1}~ & 1 \\ 
readelf & 324 & 4983 & 34s & 4 & 4875 & 13088s & 3 & 4763 & 5978s & 0 & 5131 & 299s & 5 & 4144 & G & 0 & 3760 & G & 0 \\ 
nm & 324 & 1876 & 2s & 0 & 1872 & 9869s & 0 & 1891 & 4521s & 0 & 1838 & 65s & 0 & 1424 & G & 0 & 1103 & G & 0 \\ 
tiffsplit & 858 & 1721 & 4s & 6 & 1712 & 17187s & 4 & 1697 & 8743s & 6 & 935 & 19s & 4 & 1155 & G & 2 & 1134 & G & 0 \\ 
tiff2bw & 858 & 1882 & 6s & 8 & 1829 & 13096s & 4 & 1861 & 7278s & 8 & 1263 & 13s & 1 & 1533 & G & 0 & 1435 & G & 0 \\ 
objdump & 324 & 3629 & 25s & 0 & 3667 & 19014s & 0 & 3731 & 9879s & 0 & 2422 & 1143s & 0 & 2624 & G & 1 & 2275 & G & 0 \\ 
tcpdump & 451 & 11594 & 59s & 0 & 10465 & 19800s & 0 & 10827 & 7126s & 0 & 7355 & 2077s & 1 & 5948 & G & 0 & 4499 & G & 0 \\ 
mujs & 451 & 4153 & 0s & 0 & 3973 & 15638s & 0 & 4027 & 6055s & 0 & 2421 & 8605s & 0 & 2590 & G & 1 & 2418 & G & 0 \\ 
size & 324 & 1752 & 4s & 0 & 1733 & 7343s & 0 & 1747 & 3455s & 0 & 1672 & 128s & 0 & 1331 & G & 0 & 1225 & G & 0 \\ 
infotocap & 432 & 1824 & 131s & 7 & 1782 & 33321s & 6 & 1679 & 34827s & 5 & 895 & 160s & 0 & 1164 & G & 0 & 1308 & G & 0 \\ 
imginfo & 432 & 1625 & 2s & 0 & 1494 & 26077s & 0 & 1611 & 24070s & 0 & 1307 & 43s & 0 & 1413 & G & 0 & 1105 & G & 0 \\

\midrule
\textbf{Total} & ~ & \textbf{39422} & & \textbf{26} & 37895 & & 17 & 38283 & & 20 & 27698 & & 11 & 26724 & & 5 & 22971 & & 1 \\


\bottomrule
\end{tabular}
\begin{tablenotes}[para]
\small
\item[1]G: A GPU is required to train the model.
\end{tablenotes}
\end{threeparttable}
}

\label{seeds_1000}
\end{table*}

\subsection{\tool{} vs. Byte-Scheduling Fuzzers} \label{subsec:eval_cmp_byte}
In this section, we compare \tool{} with five state-of-the-art byte-scheduling fuzzers. We analyze their performance on three sets of initial seeds with different lengths, observing their effectiveness in edge coverage, bug detection, and analysis time.

\subsubsection{Compared Fuzzers}
There are three main types of existing methods for identifying constraint-related bytes in seed files, including inference-based fuzzers, neural network-based fuzzers, and taint-based fuzzers. In order to compare the performance of \tool{} with these three category fuzzer in different scenarios, we designed a diverse set of experiments to analyze the pros and cons of each fuzzer. 
Specifically, we compared \tool{} with five state-of-the-art fuzzers Greyone~\cite{greyone}, ProFuzzer~\cite{profuzzer}, NEUZZ~\cite{neuzz}, PreFuzz~\cite{prefuzz} and Angora\cite{angora}.

\noindent\textbf{Inference-based fuzzer.}
Inference-based fuzzers mutate all bytes of a seed before scheduling mutation, and determine the importance of each byte based on its execution, such as Greyone~\cite{greyone} and ProFuzzer~\cite{profuzzer}. Greyone iterates over each byte in the input, observing whether the state of a CMP instruction in the program changes. If it does, then the byte is related to the target CMP. ProFuzzer iterates over all 256 values at each byte and determines the type of each byte (e.g., size and offset) based on changes in the path length. However, since these two fuzzers are not open-sourced, we implement their core ideas based on AFL++. For GreyOne, we mutate each byte of the seed and observe whether the number of CMPs with unexplored branches changes. If it does, we consider the byte to be CMP-related. For ProFuzzer, we mutate each byte of the seed four times and observe whether the execution path changes. If it does, we consider the byte to be control flow-related. Finally, we randomly mutate the identified CMP-related bytes or control flow-related bytes.

\noindent\textbf{Neural network-based fuzzer.}
Neural network-based fuzzers refer to the ones that use neural networks to roughly identify the constraint-related bytes of a seed, such as NEUZZ~\cite{neuzz} and PreFuzz~\cite{prefuzz}. NEUZZ identifies bytes related to program branches using deep learning models. PreFuzz optimizes the edge and byte selection mechanism of NEUZZ.

\noindent\textbf{Taint-based fuzzer.}
Taint-based fuzzers refer to fuzzers that use taint analysis to analyze the bytes related to path constraints, such as Angora~\cite{angora}. Angora uses byte-level taint analysis to track which input bytes are related to path constraints.

\subsubsection{Experiment Setup}
The recognition of constraint-related bytes by inference-based fuzzers, neural network-based fuzzers, and taint-based fuzzers is significantly influenced by the length of the seed. The longer the seed, the longer the analysis time required by these methods. Seeds with long length may fail the execution of existing fuzzers. However, \tool{} does not need to analyze each byte of the seed one by one, nor does it require building a neural network model. Therefore, the analysis time of \tool{} is not significantly affected by the length of the seed. To analyze the performance of each fuzzer on initial seeds of different lengths, we divide the initial seeds into three groups based on their length and conduct experiments on each of them separately. The length related groups are \texttt{ALL SEEDS}, \texttt{SEEDS<10000} (seeds that have less than 10,000 bytes), and \texttt{SEEDS<1000} (seeds that have less than 1,000 bytes). We analyze the analysis time, edge coverage, and bug discovery of each fuzzer in the three experimental environments. All experiments run for 24 hours and repeat 5 times.

\subsubsection{Edge Coverage and Analysis Time}
Tables \ref{all_seeds}, \ref{seeds_10000} and \ref{seeds_1000} show the edge coverage and analysis time achieved by six fuzzers on three sets of different initial seeds. From Table \ref{all_seeds}, with all the initial seeds provided, \tool{} discovers a total of 59681 edges, which are 4170 more edges than the second best fuzzer, ProFuzzer. Notably, NEUZZ and PreFuzz only discover 20213 and 23470 edges (about 1/3 edges discovered by \tool{}) respectively on 16 programs. On programs like \texttt{lame}, NEUZZ and PreFuzz fail to run the fuzzer because of long seed lengths, 
resulting in a large model that causes an OOM error during training.
As for analysis time, \tool{} consistently requires less time to analyze all programs compared to other fuzzers. For example, \tool{} uses less than 1000 seconds to analyze 13 programs, while GreyOne and ProFuzzer require more than 10000 seconds for most programs. This is because GreyOne and ProFuzzer have to analyze each byte of the seed before mutation, which means that the analysis time is proportional to the seed length. Angora requires less than 1000 seconds to analyze 7 programs, indicating that the \tool{}'s analysis time is still superior to the traditional taint analysis.
The average analysis times for \tool{}, GreyOne, ProFuzzer, and Angora are  640.37 seconds, 54196.06 seconds, 36174.81 seconds, and 2567.31 seconds, respectively. The average analysis time required by \tool{} is only about 2\% of Greyone's.
NEUZZ and PreFuzz require an additional GPU for analysis, which consumes a significant amount of resources.

From Table \ref{seeds_10000} and \ref{seeds_1000}, as the length of the seed decreases, the analysis time of all fuzzers decreases, particularly for GreyOne and ProFuzzer.
For example, with \texttt{ALL SEEDS}, ProFuzzer takes 81747 seconds to analyze \texttt{nm}, but with \texttt{SEED < 1000}, the analysis time decreases to 4521 seconds.
Notably, \tool{} exhibits the most stable performance among the three experiments and consistently achieves the highest edge coverage on most programs.
Finally, \tool{} discovers the most total edges across all three experiments.

\subsubsection{Unique Bugs}
Tables \ref{all_seeds}, \ref{seeds_10000} and \ref{seeds_1000} show the number of unique bugs discovered by each fuzzer on three sets of initial seeds.
In all cases, \tool{} is able to discover more unique bugs than other fuzzers. On \texttt{ALL SEEDS}, \tool{} finds 103 unique bugs, which is 19 more than the best performing comparison fuzzer, Angora. On \texttt{SEEDS<10000}, \tool{} discovers 81 unique bugs, which is 11 more than the second-ranked fuzzer, Angora. On \texttt{SEEDS<1000}, \tool{} discovers 40 unique bugs, which is 3 more than the best performing comparison fuzzer, ProFuzzer.
Notably, we find that a fuzzer's performance in terms of edge coverage and bug discovery is not always consistent. For example, although Angora does not perform as well as GreyOne and ProFuzzer in terms of edge coverage, it still discovers more bugs. \tool{} achieves the best performance in both edge coverage and bug discovery.

Further analysis shows that \tool{} finds 36 bugs that are not found by any other fuzzers. Among the 36 bugs, 18 of them come from the code that is explored but without crashing it by the others while the remaining 18 of them come from completely new code regions.

\subsubsection{Bug Intersection Analysis}
We compute set differences and intersections for bugs in Table~\ref{all_seeds}. The results are shown in Table~\ref{bug_insertion}. \tool{} discovers 64, 52, 75, 102, and 96 bugs that are not found by GreyOne, ProFuzzer, Angora, NEUZZ, and PreFuzz, respectively. GreyOne, ProFuzzer, NEUZZ, and PreFuzz detect fewer than 10 bugs that are not found by \tool{}, but Angora detects 54 bugs that \tool{} misses. Further analysis reveals that Angora performs well on the program \texttt{nm} by finding 39 bugs that \tool{} misses. However, \tool{} discovers 479 more new edges than Angora.
We believe that Angora's lower efficiency in discovering new edges is due to its constraint analysis method, which explores constraints one by one. Angora's constraint prioritization strategy might contribute to its ability to finding more bugs, but this could be coincidental. \tool{}'s limited exploitation capability may also be a reason for its mediocre performance on some programs.

\begin{table}[htbp]
\centering
\caption{Set differences and intersections for bugs in Table~\ref{all_seeds}.}
\resizebox{\linewidth}{!}{
\begin{threeparttable}
\begin{tabular}{@{\extracolsep{4pt}}lccccc@{}} 
\toprule
\textbf{} &   \textbf{GreyOne} & \textbf{ProFuzzer} & \textbf{Angora} & \textbf{NEUZZ} & \textbf{PreFuzz}\\
\midrule
\tool{} \& *\tnote{1} & 39 & 51 & 30 & 1 & 7\\
\tool{} - *\tnote{2} & 64 &52 &73 &102 &96\\
* - \tool{}\tnote{3} & 8 &8 &54 &1 &2\\

\bottomrule
\end{tabular}
\begin{tablenotes}[para]
\item[1] \tool{} \& *: the intersection of bugs found by both \tool{} and the other fuzzers.
\item[2] \tool{} - *: the number of bugs found by \tool{} but not by the other fuzzers.
\item[3] * - \tool{}: the number of bugs found by other fuzzers but not by \tool{}.
\end{tablenotes}
\end{threeparttable}
}
\label{bug_insertion}
\end{table}


\subsection{\tool{} vs. Commonly Used Fuzzers}
In this section, we compare \tool{} with some commonly used fuzzers to further demonstrate its superior performance. 
To ensure fair results, we conduct our experiments on two third-party testing platforms, UNIFUZZ~\cite{unifuzz} and MAGMA~\cite{magma}.
All experiments run for 24 hours and repeat 10 times to avoid the randomness of fuzzing~\cite{bohme2022reliability}.

\subsubsection{Compared Fuzzers}
We compare \tool{} with the state-of-the-art CGFs, including AFL~\cite{afl}, AFL++~\cite{aflpp}, AFLFast~\cite{aflfast}, FairFuzz~\cite{fairfuzz}, MOPT~\cite{MOPT}, and EMS~\cite{ems}.
In this section, we use the AFL, AFLFast, MOPT, and FairFuzz implemented within AFL++ for our experiments.
AFL is a popular fuzzer that is the basis of many other recent fuzzers~\cite{aflfast, fairfuzz, MOPT, Yue2020EcoFuzz, ems, neuzz}. 
AFL++ is a framework that integrates many state-of-the-art fuzzers into AFL, such as MOPT, AFLfast and Redqueen.
In our evaluation, We enable AFL++'s Redqueen mutator and MOPT mutator, which is the best-performing configuration of AFL++.
AFLFast models the fuzzing process as a Markov chain and prioritizes low-frequency paths to achieve better code coverage. FairFuzz proposes a novel mutation mask to guide fuzzing towards rare branches. MOPT utilizes Particle Swarm Optimization to optimize the schedule of mutation operators based on fuzzing history. 
For MOPT and MOPT related fuzzers, we set $-L = 5$. 

\subsubsection{Edge Coverage}
\begin{figure}[!t]
\centering
\includegraphics[width=0.9\linewidth]{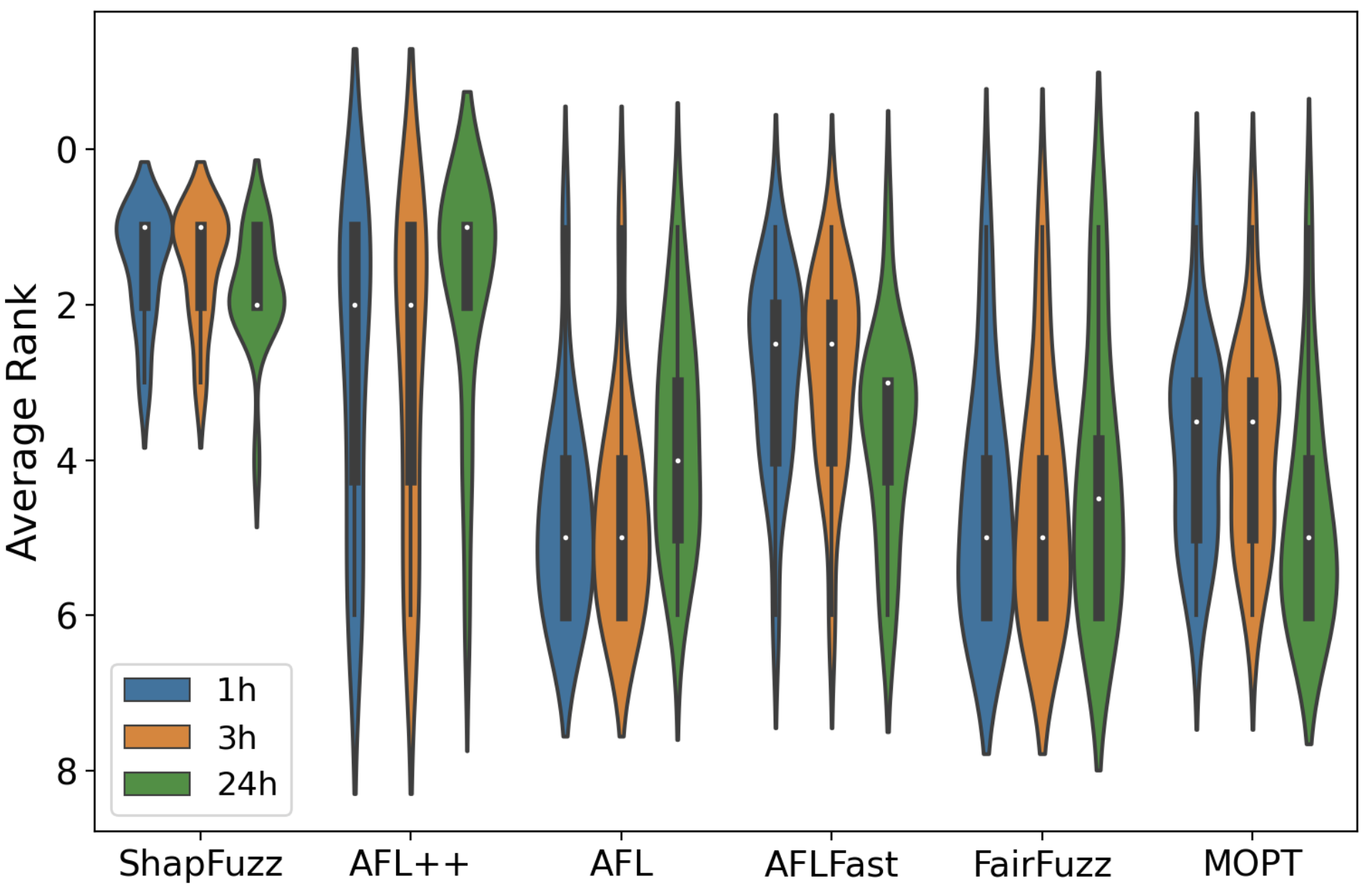}
\caption{The violin plot of average edge ranks across 10 trials of different fuzzers after running for 1/3/24 hours. 
}
\label{average_ranks}
\end{figure}

To evaluate the efficiency of edge discovery, we evaluate fuzzers by using the ranks of edge discovery after fuzzing 1/3/24 hours. The lower the rank is, the better the fuzzer performs. As shown in Figure~\ref{average_ranks}, the performance of most fuzzers at different running hours is not significantly different.
Yet, Figure~\ref{average_ranks} indicates that \tool{} achieves better code coverage on most programs than other fuzzers.
After fuzzing for 1 and 3 hours, the median rank of \tool{} is 1.5 and 1.5, respectively, which indicates that the fuzzing efficiency of \tool{} in a short period of time is higher.
After fuzzing 24 hours, the median rank of \tool{} is 1.81. 
This indicates that \tool{}'s strategy is more efficient than MOPT whose the median rank is 4.75. 

\begin{table}[!t]
\centering
\caption{The number of unique bugs discovered by different fuzzers (on UNIFUZZ).}
\resizebox{\linewidth}{!}{
\begin{threeparttable}
\begin{tabular}{@{\extracolsep{4pt}}cccccccccccc@{}} 
\toprule


\textbf{Programs} & \textbf{\tool{}} & \textbf{AFL++} & \textbf{MOPT} & \textbf{AFL} & \textbf{AFLFast} & \textbf{FairFuzz}\\
\midrule
tiff2pdf           &      1 &           2 &   1 &      0 &   0 &   0 \\
lame               &      4 &            4 &   4 &      5 &   4 &   4 \\
readelf         &      6 &            4 &   4 &      3 &   5 &   5 \\
exiv2              &     18 &           23 &   3 &     15 &  15 &  14 \\
flvmeta            &      2 &            2 &   2 &      2 &   2 &   2 \\
nm              &     35 &           31 &  21 &     22 &  22 &  30 \\
tiffsplit          &      7 &            9 &   7 &      5 &   7 &   5 \\
tiff2bw            &     10 &            6 &   5 &      5 &   5 &   7 \\
objdump         &     29 &           11 &   8 &     12 &  10 &   8 \\
pdftotext          &     27 &           24 &  20 &     25 &  24 &  23 \\
mp42aac            &      0 &            5 &   0 &      0 &   0 &   0 \\
tcpdump            &      2 &            1 &   0 &    2 &   1 &   2 \\
infotocap          &    11 &            9 &   9 &      5 &   9 &   9 \\
imginfo            &      0 &            1 &   0 &     2 &   0 &   0 \\

\midrule
\textbf{Total}  &  \textbf{152} & 132  & 84  & 103 & 104 & 109\\

\bottomrule
\end{tabular}
\end{threeparttable}
}

\label{unifuzz_bug}
\end{table}

\subsubsection{Bug Discovery on UNIFUZZ}
\begin{sloppypar}
We analyze the number of unique bugs triggered by a fuzzer based on ASAN. Following the guidance of UNIFUZZ~\cite{unifuzz}, we select the top three stack frames to de-duplicate bugs. 
Table~\ref{unifuzz_bug} shows the number of unique bugs triggered by each fuzzer. \tool{}, MOPT, AFLFast, FairFuzz, AFL, and AFL++ trigger 152, 84, 103, 104, 109 and 132 bugs, respectively. 
\tool{} finds the most bugs in 16 programs, with 20 more bugs than AFL++. 
Specifically, \tool{} performs the best on 8 programs, including \texttt{nm}, \texttt{readelf}, \texttt{flvmeta}, \texttt{tiff2bw}, \texttt{objdump}, \texttt{pdftotext}, \texttt{tcpdump} and \texttt{infotocap}. 
Surprisingly, MOPT and AFLFast do not perform the best on any program. 
AFL++, AFL and FairFuzz only perform the best on 5, 3 and 1 programs, respectively.
In summary, \tool{} performs the best on most of the programs, demonstrating the effectiveness of \tool{}. 

Furthermore, we use \tool{} to test some of the latest versions of programs and successfully found several previously unknown bugs. The results are shown in the Table~\ref{latestVersion}.
We find 11 bugs in 6 programs. Particularly, two bugs have been confirmed by developers as previously undiscovered bugs, and one bug has been fixed in the beta version.
The four bugs in \texttt{jhead} have been fixed in the new version before receiving a response.
The remaining 4 bugs are still being confirmed.
\end{sloppypar}

\begin{table}[!t]
\centering
\caption{New bugs discovered by \tool{} on the latest versions of programs.}
\resizebox{0.9\linewidth}{!}{
\begin{threeparttable}
\begin{tabular}{@{\extracolsep{4pt}}lccc@{}} 
\toprule
\textbf{Programs} & \textbf{Known Vulnerabilities}  & \textbf{Unknown Vulnerabilities} & \textbf{Sum}\\
\midrule
infotocap & 0 & 1 & 1\\
imginfo & 0 & 1 & 1\\
mp42aac & (submitted)\tnote{1} & (submitted) & 2\\
jhead & 4 & 0 & 4\\
cflow & (submitted) & (submitted) & 2\\
mujs & 1 & 0 & 1\\
\hline
Total & & & 11 \\
\bottomrule
\end{tabular}
\begin{tablenotes}[para]
(submitted): We have submitted the bugs and have no response yet. 
\end{tablenotes}
\end{threeparttable}
}
\label{latestVersion}
\end{table}

\subsubsection{Overhead}
\begin{figure}[!t]
\centering
\includegraphics[width=0.85\linewidth]{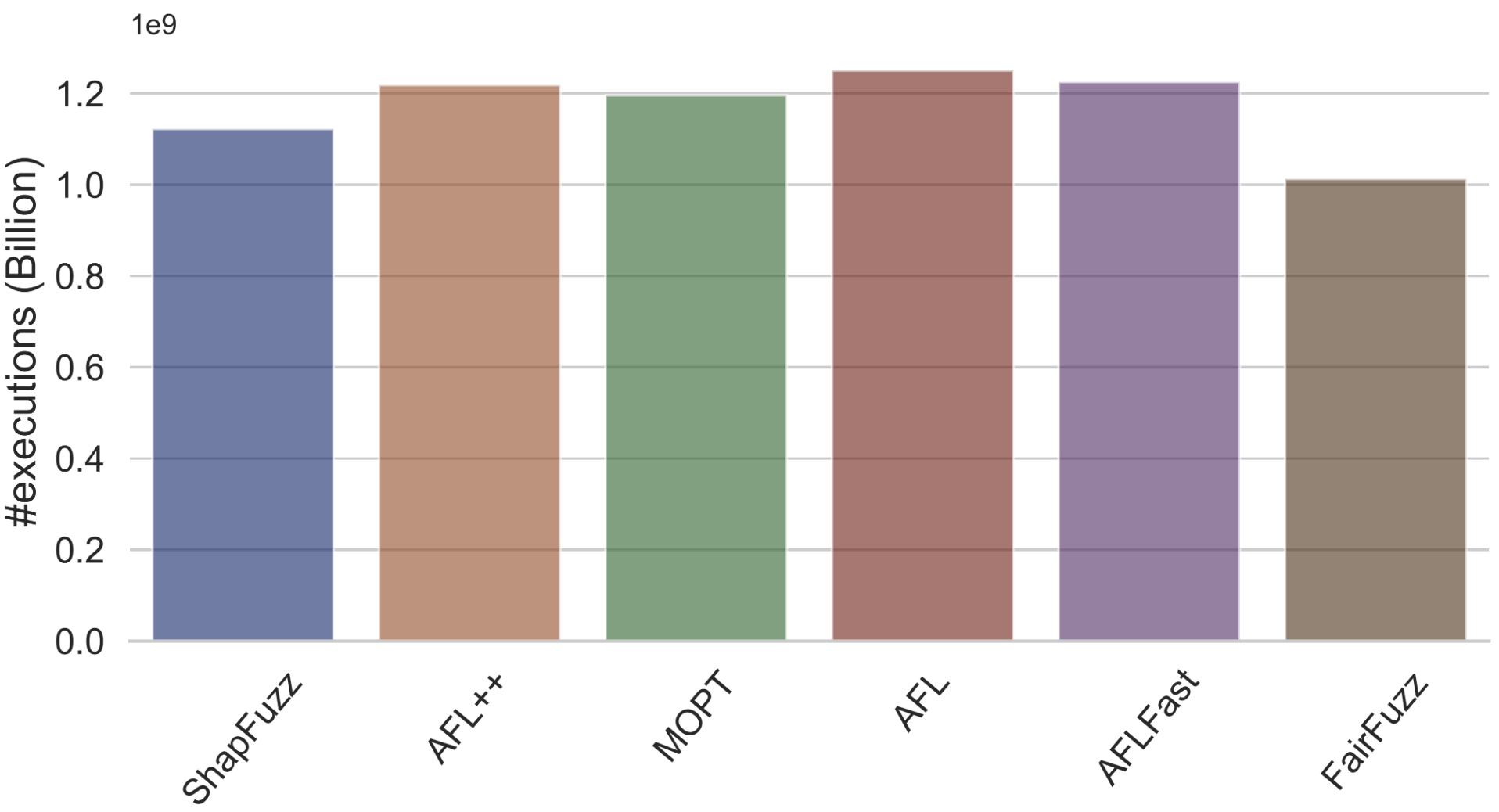}
\caption{The total throughput of different fuzzers running for 24 hours.}
\label{throughput}
\end{figure}

\tool{} adds some extra steps to AFL++, including tracing mutators, recovering mutators, and calculating the Shapley Values of bytes. Thus, to quantify the influence of these steps on fuzzing throughput (\ie execution speed), we measure the total number of executions for each fuzzer. Figure~\ref{throughput} shows the total number of executions of all 16 programs that fuzzers run after 24 hours.
The results show that the throughput of \tool{} does not drop significantly comparing to other fuzzers.
After fuzzing 24 hours, the throughput of \tool{} is lower than that of MOPT, AFL++, AFL and AFLFast by 6.16\%, 7.91\%, 10.25\%, 8.38\%, respectively.
However, the total throughput of \tool{} is higher than FairFuzz by 9.75\%.
Surprisingly, on programs \texttt{tiffsplit} and \texttt{exiv2}, the throughput of \tool{} is higher than AFL++, MOPT, AFL and FairFuzz. 
Therefore, our \tool{} does not introduce much overhead.

\subsubsection{Evaluation on MAGMA}

\begin{table}[!t]
\centering
\caption{Target programs from MAGMA.}\label{programsMAGMA}
\resizebox{0.9\linewidth}{!}{
\begin{threeparttable}
\begin{tabular}{llll}
    \toprule
    \textbf{Projects} & \textbf{Target Programs} & \textbf{Input format}\\ \midrule
    \textbf{libpng 1.6.38} & libpng\_read\_fuzzer & PNG\\
    \midrule
    \textbf{libtiff 4.1.0} & read\_rgba\_fuzzer, tiffcp & TIFF\\
    \midrule
    \multirow{2}{*}{\textbf{libxml2 2.9.10}} & \multirow{2}{*}{\makecell[c]{libxml2\_xml\_read\_memory\_fuzzer, \\ xmllint}} & \multirow{2}{*}{XML}\\
    &&\\
    \midrule
    \textbf{poppler 0.88.0} & pdf\_fuzzer, pdfimages & PDF\\
    \midrule
    \multirow{3}{*}{\textbf{openssl 3.0.0}} & \multirow{3}{*}{\makecell[c]{asn1, asn1parse, bignum, server, \\ client, x509}} & \multirow{3}{*}{Binary blobs}\\
    &&\\
    &&\\
    \midrule
    \textbf{sqlite3 3.32.0} & sqlite3\_fuzz & SQL queries\\
    \midrule
    \multirow{2}{*}{\textbf{php 8.8.0}} & \multirow{2}{*}{\makecell[c]{exif, json, parser, \\ unserialize}} & \multirow{2}{*}{Various}\\
    &&\\
    \midrule
    \textbf{libsndfile } & sndfile\_fuzzer & WAV\\
    \midrule
    \textbf{lua } & lua & LUA\\
    \bottomrule
\end{tabular}
\end{threeparttable}

}
\end{table}

To further evaluate the bug-finding performance of \tool{}, we conduct experiments on MAGMA.
MAGMA is a ground-truth fuzzing benchmark that enables accurate and consistent fuzzing evaluation. MAGMA provides several benchmarks with real-world bugs, and it can perform a realistic evaluation of fuzzers.
We integrate our \tool{} into MAGMA, and compare them to 5 widely-used mutation-based fuzzers (\ie AFL, AFLFast, FairFuzz, MOPT and AFL++). AFL++ is the AFL++ that enables both MOPT and Redqueen. Due to configure issues of MAGMA, we use the original implementations of fuzzers rather than the implementations based on AFL++. Table~\ref{programsMAGMA} shows the target programs used in MAGMA.

\noindent\textbf{Bug Discovery.}
Table~\ref{magma_unique_bugs} shows the number of CVEs triggered by each fuzzer on average. 
It shows that \tool{} performs the best on MAGMA and it exposes the most bugs on 7 of 9 programs.  
Specifically, \tool{} performs the best on \texttt{libsndfile}, \texttt{libxml2}, \texttt{lua}, \texttt{openssl}, \texttt{php}, \texttt{poppler} and \texttt{sqlite3}, exposing 7.00, 4.22, 1.00, 4.00, 3.00, 5.56 and 4.56 bugs on average, respectively.
Moreover, \tool{} performs significantly better than all the other fuzzers on \texttt{openssl}, \texttt{poppler} and \texttt{sqlite3}.
On \texttt{splite3}, \tool{} discovers almost two times the number of bugs compared to AFL++. 
Surprisingly, AFL++ only exposes the most bugs on 2 programs.

\begin{table}[h]
\centering
\caption{The average number of bugs found by different fuzzers (on MAGMA).}
\resizebox{\linewidth}{!}{
\begin{threeparttable}
\begin{tabular}{@{\extracolsep{4pt}}cccccccccccc@{}} 
\toprule

\textbf{Programs} & \textbf{\tool{}} &  \textbf{AFL++} & \textbf{AFL} & \textbf{AFLFast} & \textbf{FairFuzz} & \textbf{MOPT}\\
\midrule
libpng    & 1.44  &   \textbf{2.22} & 1.89 & 1.00 & 1.56 & 1.56 \\
libsndfile  & \textbf{7.00}  & 6.89 & 2.89 & 2.33 & 1.44 & 4.78 \\
libtiff   & 5.89   &  \textbf{6.22} & 5.44 & 4.22 & 4.67 & 5.67 \\
libxml2   & \textbf{4.22}   &  4.00 & 1.00 & 1.89 & 1.00 & 2.78 \\
lua       & \textbf{1.00}   &  0.56 & \textbf{1.00} & \textbf{1.00} & 0.89 & \textbf{1.00} \\
openssl    & \textbf{4.00}  &  2.67 & 3.11 & 3.33 & 2.44 & 3.00 \\
php        & \textbf{3.00}  &  2.00 & \textbf{3.00} & \textbf{3.00} & 2.56 & \textbf{3.00} \\
poppler    & \textbf{5.56}  &  4.33 & 3.00 & 3.44 & 2.22 & 3.44 \\
sqlite3    & \textbf{4.56}  &  2.56 & 0.89 & 1.78 & 0.89 & 0.67 \\
\midrule
\#The Best\tnote{1} & \textbf{7}  & 2 & 2 & 2 & 0 & 2\\
\bottomrule
\end{tabular}
\begin{tablenotes}[para]
\item[1]\#The Best: the number of programs that a fuzzer performs the best.
\end{tablenotes}
\end{threeparttable}
}
\label{magma_unique_bugs}
\end{table}

\noindent\textbf{Time to Bug.}
We analyze the first time that each bug is found and the results are shown in the Table~\ref{TTB}. 
In 10 trials, a total of 48 bugs are found by these 7 fuzzers. Among them, \tool{} performs the best, with a total of 46 discovered bugs. 
Additionally, among the bugs discovered by \tool{}, 7 bugs cannot be exposed by AFL++, MOPT, FairFuzz, AFL and AFLFast.
As for TTB, \tool{} has the least time to bug (TTB) on 29 CVEs, which is the best performance among fuzzers.
For example, for 25 bugs, \tool{} discovers them 2$\times$ faster than AFL++.
Note that, \texttt{CVE-2017-2518} and \texttt{CVE-2019-19317}, \texttt{CVE-2017-9776}, \texttt{CVE-2013-7443}, \texttt{CVE-2019-19880}, \texttt{PDF008} and \texttt{CVE-2019-19646} can only be exposed by \tool{}.
Due to the advantage of solving path constraints, \tool{} exposes bugs in less time than others.

\begin{table}[!t]
\centering
\caption{The time to bug (TTB) of fuzzers (on MAGMA). \tool{} performs the best in discovering bugs.}
\resizebox{\linewidth}{!}{
\begin{threeparttable}
\begin{tabular}{@{\extracolsep{4pt}}crrrrrrrrrrr@{}}
\toprule
\rotatebox{0}{\textbf{Vulnerabilities}} & \rotatebox{60}{\textbf{\tool{}}}   & \rotatebox{60}{\textbf{AFL++}} & \rotatebox{60}{\textbf{MOPT}} & \rotatebox{60}{\textbf{FairFuzz}}  &  \rotatebox{60}{\textbf{AFL}} & \rotatebox{60}{\textbf{AFLFast}} \\
\midrule

\rowcolor{gray!20} CVE-2015-8472 & 17s & 23s & \textbf{15s} & \textbf{15s} & \textbf{15s} & \textbf{15s} \\ 
CVE-2016-1762 & 47s & 1m & \textbf{15s} & 20s & 16s & 18s \\
\rowcolor{gray!20} CVE-2018-13988 & 1m & 1m & \textbf{52s} & 2m & 1m & 1m \\
CVE-2016-2109 & 3m & 4m & \textbf{1m} & 2m & \textbf{1m} & 2m \\
\rowcolor{gray!20} CVE-2016-6309 & 4m & 5m & \textbf{1m} & 2m & \textbf{1m} & \textbf{1m} \\
CVE-2016-10270 & \textbf{23s} & 48s & 19m & 3m & 23m & 11m \\
\rowcolor{gray!20} CVE-2016-3658 & \textbf{4m} & 17m & 1h & 25m & 1h & 1h \\
CVE-2018-14883 & 18m & 1h & 3m & 2h & \textbf{2m} & 3m \\
\rowcolor{gray!20} CVE-2017-6892 & \textbf{13m} & 32m & 33m & 3h & \textbf{13m} & 1h \\
SND017 & \textbf{4m} & 5m & 25m & 21h & 38m & 5m \\
\rowcolor{gray!20} CVE-2017-11613 & \textbf{9m} & 1h & 1h & 15h & 6h & 4h \\
CVE-2019-11034 & 6h & 15h & \textbf{1m} & 8m & 2m & \textbf{1m} \\
\rowcolor{gray!20} PDF010 & 3h & 1h & 7h & 16h & 10h & \textbf{2m} \\
CVE-2019-7663 & \textbf{3h} & 8h & 4h & 9h & 16h & 6h \\
\rowcolor{gray!20} CVE-2019-20218 & \textbf{49m} & 2h & 16h & 11h & 9h & 9h \\
CVE-2019-9638 & \textbf{41m} & 17h & 7h & 17h & 4h & 44m \\
\rowcolor{gray!20} CVE-2015-8317 & \textbf{22m} & 1h & 3h & - & - & 12h \\
CVE-2019-7310 & \textbf{7h} & 17h & 9h & 15h & \textbf{7h} & 8h \\
\rowcolor{gray!20} CVE-2020-15945 & 5h & \textbf{4h} & 14h & 15h & 13h & 16h \\
SND020 & \textbf{16m} & 19m & 7h & 21h & - & - \\
\rowcolor{gray!20} CVE-2015-3414 & \textbf{1h} & 4h & 17h & 16h & 23h & 15h \\
CVE-2016-10269 & 2h & \textbf{1h} & 15h & 22h & 15h & - \\
\rowcolor{gray!20} CVE-2011-2696 & \textbf{18m} & 22m & 17h & - & 22h & 23h \\
CVE-2017-8363 & \textbf{13m} & 28m & 18h & 22h & - & - \\
\rowcolor{gray!20} CVE-2017-8363 & \textbf{27m} & 32m & 18h & 22h & - & - \\
CVE-2017-9047 & \textbf{33m} & 6h & 12h & - & - & - \\
\rowcolor{gray!20} CVE-2017-8361 & \textbf{14m} & 3h & 18h & 22h & - & - \\
CVE-2016-6302 & 9h & 23h & 16h & - & \textbf{3h} & 6h \\
\rowcolor{gray!20} CVE-2017-7375 & 6h & \textbf{1h} & - & - & - & - \\
CVE-2018-10768 & \textbf{5h} & \textbf{5h} & 23h & - & - & 23h \\
\rowcolor{gray!20} CVE-2016-5314 & 19h & 19h & 11h & 16h & \textbf{4h} & - \\
CVE-2013-6954 & 13h & 19h & 16h & 12h & \textbf{11h} & - \\
\rowcolor{gray!20} CVE-2019-19926 & \textbf{5h} & 19h & 17h & - & 21h & 20h \\
PNG006 & - & \textbf{6m} & - & - & - & - \\
\rowcolor{gray!20} CVE-2016-2108 & \textbf{3h} & 16h & 22h & - & - & - \\
CVE-2016-10269 & 23h & \textbf{2h} & 22h & - & - & - \\
\rowcolor{gray!20} CVE-2017-9865 & \textbf{8h} & 12h & - & - & - & - \\
CVE-2015-8784 & 22h & 21h & 20h & \textbf{18h} & - & 20h \\
\rowcolor{gray!20} CVE-2016-1836 & \textbf{22h} & 23h & - & - & - & - \\
CVE-2017-14617 & - & \textbf{21h} & - & - & - & - \\
\rowcolor{gray!20} CVE-2017-3735 & 20h & - & \textbf{5h} & 15h & 22h & 10h \\
CVE-2017-9776 & \textbf{19h} & - & - & - & - & - \\
\rowcolor{gray!20} CVE-2013-7443 & \textbf{17h} & - & - & - & - & - \\
CVE-2019-19880 & \textbf{18h} & - & - & - & - & - \\
\rowcolor{gray!20} PDF008 & \textbf{20h} & - & - & - & - & - \\
CVE-2019-19646 & \textbf{22h} & - & - & - & - & - \\
\rowcolor{gray!20} CVE-2017-2518 & \textbf{22h} & - & - & - & - & - \\
CVE-2019-19317 & \textbf{22h} & - & - & - & - & - \\
\rowcolor{gray!50}\#The Fastest\tnote{1} & \textbf{29} & 7 & 7 & 2 & 9 & 4 \\

\bottomrule
\end{tabular}
\begin{tablenotes}[para]
\item[1]\#The Fastest: The number of vulnerabilities that are discovered the fastest by the fuzzer.
\end{tablenotes}
\end{threeparttable}
}
\label{TTB}
\end{table}

\begin{figure}[htbp]
\centering
\includegraphics[width=0.9\linewidth]{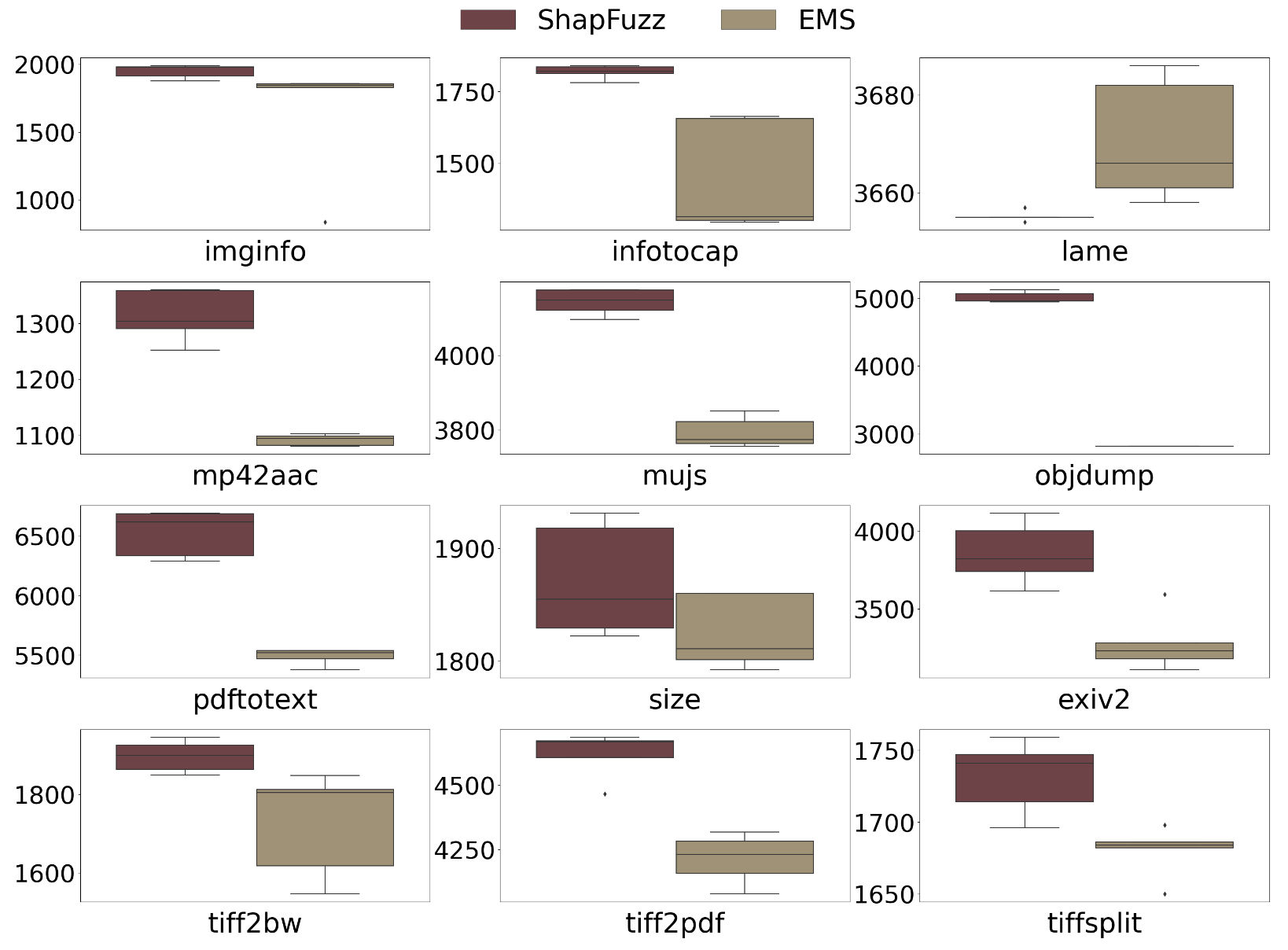}
\caption{The boxplot of edge coverage for \tool{} and EMS. Y-axis: the number of discovered edges.}
\label{tool_vs_ems}
\end{figure}

\subsection{Comparison with Byte Mutation Solution}

EMS is the latest fuzzer focusing on byte mutation, which reuses history byte values that trigger unique paths and crashes. 
Specifically, EMS finds that the fuzzing history, both within and across trials, contains valuable mutation strategies that can lead to the discovery of new paths or crashes sharing similar partial path constraints.
Because some programs will occupy tens of GB of memory when tested by EMS (implementation issues of EMS), we run experiments with part of the programs on Table~\ref{programsUNIFUZZ}. To ensure data consistency, we use \textit{afl-showmap} of AFL++ to count the number of edges for EMS and \tool{}.
The results of edge coverage discovered by \tool{} and EMS are shown in the Figure~\ref{tool_vs_ems}. 
Except for the program \texttt{lame}, \tool{} performs better than EMS.
Specifically, the median of edge coverage discovered by \tool{} is higher than EMS by 6.78\%, 27.89\%, 16.10\%, 9.01\%, 43.13\%, 16.57\%, 2.37\%, 15.51\%, 5.00\%, 9.38\% and 3.27\% on \texttt{imginfo}, \texttt{infotocap}, \texttt{mp42aac}, \texttt{mujs}, \texttt{objdump}, \texttt{pdftotext}, \texttt{size}, \texttt{exiv2}, \texttt{tiff2bw}, \texttt{tiff2pdf} and \texttt{tiffsplit}, respectively.
As to the performance of unique bugs discovered by the two fuzzers, the results are shown in Table \ref{unifuzz_bug_ems}. \tool{} triggers more bugs on 6 programs, while EMS only triggers more bugs on 2 programs. 
Our \tool{} discovers 90 bugs on these programs while EMS only discovers 39 bugs.

\begin{table}[htbp]
\centering
\caption{The number of unique bugs discovered by \tool{} and EMS (on UNIFUZZ).}
\resizebox{0.5\linewidth}{!}{
\begin{threeparttable}
\begin{tabular}{@{\extracolsep{4pt}}lcc@{}} 
\toprule
\textbf{Programs} &   \textbf{EMS} & \textbf{\tool{}}\\
\midrule
lame       &  4 &     4 \\
exiv2      & 13 &    15 \\
tiffsplit  &  8 &     6 \\
tiff2bw    &  8 &     8 \\
objdump &  1 &    15 \\
pdftotext  &  1 &    25 \\
mp42aac    &  1 &     0 \\
infotocap  &  3 &    10 \\
tiff2pdf   &  0 &     2 \\
readelf &  0 &     5 \\
\hline
Total & 39 & \textbf{90} \\

\bottomrule
\end{tabular}
\end{threeparttable}
}
\label{unifuzz_bug_ems}
\end{table}

\subsection{Ablation Study of \tool{}}

\tool{} has two main components that are calculation of Shapley values (Sections~\ref{subsec:cg_all} and \ref{sec:shap_update}) and Shapley-guided byte selection (Section~\ref{subsec:mutate}).
Since Shapley-guided byte selection cannot be performed without Shapley values, we do not compare this component. Since the process of calculating the Shapley value introduces some additional mutations, we want to analyze the impact they have. To analyze the impact of calculating Shapley values, we implement \tool{}-modeling to compare with \tool{}. For \tool{}-modeling, we remove the Shapley-guided byte selection and randomly select positions to mutate.

\begin{figure}[t!]
\centering
\includegraphics[width=0.9\linewidth]{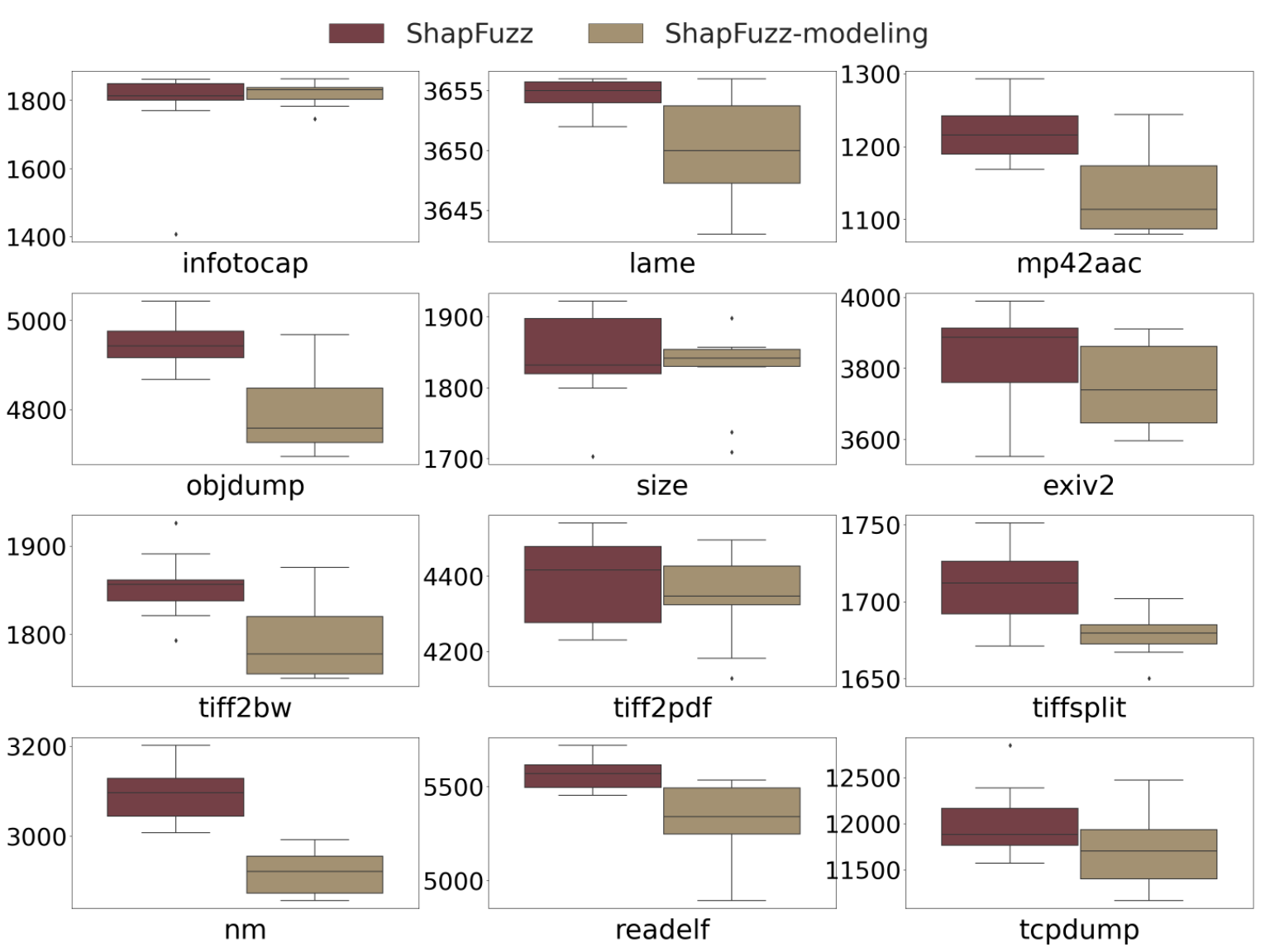}
\caption{The boxplot of edge coverage for \tool{} and
\tool{}-modeling. Y-axis: the number of discovered edges.}
\label{ablation}
\end{figure}

We evaluated two fuzzers \tool{} and \tool-modeling in UNIFUZZ for 24 hours for 5 times.
The result is shown in the Figure~\ref{ablation}.
With the help of Shapley-guided mutation, \tool{} discovers more new edges under most procedures. 
For instance, the median of \tool{} is higher than \tool-modeling on \texttt{objdump}, \texttt{nm}, \texttt{tiffsplit}, \texttt{lame}, \texttt{exiv2}, \texttt{tiff2bw}, \texttt{tcpdum} and \texttt{mp42aac}.
It illustrates that Shapley-guided mutation can indeed generate a more useful mutation possibility distribution for seeds.

\subsection{Comparison with Inference-based Fuzzers} \label{subsec:cmp_infer}
\balance
In this section, we compare \tool{} with inference-based fuzzers, including GreyOne~\cite{greyone} and ProFuzzer~\cite{profuzzer}, in terms of the effectiveness of inferred bytes. 



\noindent\textbf{Experiment Setup.}
We randomly select 10 target programs and several initial seeds for experiments from UNIFUZZ.
To compare the efficiency gap between the Shapley bytes identified by \tool{} and the CMP-related bytes identified by GreyOne and ProFuzzer, we analyze each initial seed separately, and the specific experimental steps are as follows. 
For an initial seed, we use \tool{}, Greyone and ProFuzzer to infer the Shapley bytes or CMP-related bytes and mutate these bytes randomly for 3 hours, respectively. 
Then, we count the additional analysis time required by each fuzzer and the number of new edges they discover. In particular, for the convenience of comparison, we do not update the seed queue but continuously mutate the selected initial seed.


\noindent\textbf{Comparison Metrics.} 
We use the following two new metrics for fine-grained analysis of the above fuzzers:
\begin{itemize}
  \item [1)] 
  \textit{The proportion of useful bytes}: If a byte is inferred and finds a new edge in the subsequent mutation process, it is called a useful byte. \textit{\textbf{For \tool{}, we do not count the first time when an inferred byte is retained.}} We record the proportion of useful bytes in all inferred bytes.
  \item [2)] 
  \textit{The average number of useful mutations}: If the mutation of inferred bytes finds a new edge, this mutation is a useful mutation.
  We use this metric to describe the usefulness of each byte. 
  If a byte has more mutations that discover new edges, it is more useful.
\end{itemize}

\noindent\textbf{Result.}
We only include seeds that can complete the analysis within three hours. In three hours, \tool{} can complete the analysis of all seeds, but both GreyOne and ProFuzzer have some long seeds that cannot complete the analysis. The results are shown in the Table~\ref{Verification}.
We analyze the average number of useful mutations for each fuzzer at 1,000,000 mutations. \tool{} perform the best, and its average proportion of useful bytes is up to 70.5\%. Moreover, \tool{}'s average number of useful mutations is up to 3.715. The proportion of useful bytes and average number of useful mutations of GreyOne are only 16.1\% and 0.947, while that of ProFuzzer are only 17.8\% and 0.0887.
According to the above experimental results, \tool{} can find more useful bytes within shorter analysis time.

\begin{table}[!h]
\centering
\caption{The comparison of position inference. \tool{} can infer more useful positions.}
\resizebox{\linewidth}{!}{
\begin{threeparttable}
\begin{tabular}{@{\extracolsep{4pt}}lccccccccccc@{}}
\toprule
\multirow{2}{*}{\textbf{Programs}} & \multicolumn{2}{c}{\textbf{GreyOne}} & \multicolumn{2}{c}{\textbf{ProFuzzer}} & \multicolumn{2}{c}{\textbf{\tool}}\\
\cline{2-3} \cline{4-5} \cline{6-7}

& UR\tnote{1} & AUM\tnote{2} & UR & AUM  & UR & AUM \\



\midrule

\textbf{nm} & 3\%(135/4834) & 0.11 & 16\%(109/665) & 0.32 & \textbf{67\%(55/82)} & \textbf{1.87}  \\
\textbf{tiff2bw} & 13\%(101/762) & 0.74 & 14\%(106/762) & 0.75 & \textbf{89\%(74/83)} & \textbf{5.87}\\
\textbf{tiffsplit} & 68\%(104/152) & 3.18 & 69\%(104/150) & 3.25 & \textbf{92\%(72/78)} & \textbf{6.60}\\
\textbf{objdump} & 2\%(205/13288) & 0.45 & 18\%(318/1794) & 0.68 & \textbf{60\%(179/299)} & \textbf{1.70} \\
\textbf{readelf} & 3\%(99/3798) & 0.41 & 7\%(186/2641) & 0.48 & \textbf{83\%(95/114)} & \textbf{8.70}\\
\textbf{flvmeta} & 46\%(29/63) & 1.28 & 22\%(42/187) & 0.46 & \textbf{40\%(14/35)} & \textbf{1.08}\\
\textbf{pdftotext} & 3\%(267/8384) & 0.07 & 4\%(337/8805) & 0.08 & \textbf{54\%(165/305)} & \textbf{1.11}\\
\textbf{mp42aac} & 19\%(124/651) & 0.32 & 21\%(132/614) & 0.35 & \textbf{68\%(69/102)} & \textbf{1.70}\\
\textbf{tcpdump} & 2\%(88/4075) & 2.89 & 3\%(87/2708) & 2.48 & \textbf{77\%(72/93)} & \textbf{6.36}\\
\textbf{lame3.99.5} & 1\%(169/13077) & 0.02 & 1\%(179/13060) & 0.02 & \textbf{73\%(74/102)} & \textbf{2.16}\\
\midrule
\textbf{Average} & 16\% & 0.947 & 17.5\% & 0.887 & \textbf{70.3\%} & \textbf{3.715} \\

\bottomrule
\end{tabular}
\begin{tablenotes}[para]
\item[1]UR: Useful Ratio.
\item[2]AUM: Average Useful Mutation.
\end{tablenotes}
\end{threeparttable}
}
\label{Verification}
\end{table}

\subsection{Case Study}
In the previous experiments, we demonstrate the superiority of \tool{} in terms of edge coverage and bug discovery. In this section, we explore the insights of \tool{}.
As shown in the Figure \ref{tiff2pdf}, we select a code snippet from \texttt{tiff2pdf} to illustrate why the analysis time of \tool{} is shorter and why it can discover more new edges. In \texttt{tiff2pdf}, there are more than 10 \texttt{if} statements directly related to \texttt{t2p->tiff\_samplesperpixel}, and Figure \ref{tiff2pdf} shows only a part of them. 
Assume that a successful mutation flips one of the \texttt{if} statements, and then we analyze the high Shapley-value bytes, \ie the bytes that contribute the most to the coverage. When we want to explore the remaining \texttt{if} statements related to \texttt{t2p->tiff\_samplesperpixel} in \texttt{tiff2pdf}, we can prioritize mutating the high Shapley-value bytes extracted earlier to quickly explore these \texttt{if} statements. 
However, it is difficult to infer which \texttt{if} statements share relevant bytes with each other directly. Therefore, we use the CMAB method to continuously analyze the relationships between constraints during attempts, which can help alleviate this problem. For more details, please refer to Section \ref{subsec:mutate}.
However, for GreyOne, ProFuzzer, and Angora, the relationship between bytes and \texttt{if} statements in each path of the program needs to be re-analyzed, which consumes a lot of time.
For NEUZZ and PreFuzz, they cannot accurately analyze the relationship between input bytes and \texttt{if} statements. 

\begin{figure}[tp]
    \centering
\begin{lstlisting}

void t2p_read_tiff_data(T2P* t2p, TIFF* input){
    if(t2p->tiff_samplesperpixel>4){...}
    ...
    if(t2p->tiff_samplesperpixel==0){...}
    ...
    switch(t2p->tiff_photometric){
        ...
        case PHOTOMETRIC_RGB: 
            ...
            if(t2p->tiff_samplesperpixel == 3){...}
            ...
            if(t2p->tiff_samplesperpixel > 3){...}
            ...
    }
}
\end{lstlisting}
\caption{A partial code snippet extracted from  \texttt{tiff2pdf}. \texttt{t2p->tiff\_samplesperpixel} is used multiple times at different program locations.}\label{tiff2pdf}
\end{figure}

\section{Related Work}

Coverage-guided fuzzing aims to effectively discover new code~\cite{zhu2022roadmap, aflchurn, feng2021snipuzz, truzz}. 
The most related works to byte mutation are the ones that resolve path constraints, including hybrid fuzzing, taint-based fuzzing, and other types of fuzzing.

\noindent\textbf{Hybrid Fuzzing.}
Symbolic execution uses a constraint solver to generate an input that drives program execution to a specific path.
Hybrid fuzzing utilizes both symbolic execution and fuzzing to explore program states. In hybrid fuzzing, symbolic execution is used to solve the constraints of hard-to-cover branches when fuzzing suffers from passing path constraints. 

Although hybrid fuzzing can solve path constraints efficiently, it still suffers from path explosion inherited from symbolic execution. Therefore, state-of-the-art fuzzers selectively address path constraints to avoid this problem. Driller~\cite{driller} uses selective concolic execution to explore those uncovered paths instead of all paths. Dowser~\cite{dowser} proposes solving the paths that are most likely to trigger buffer-overflow vulnerabilities. SymFuzz~\cite{SymFuzz} leverages symbolic execution to detect dependencies among the bit positions of an input for a program. QSYM~\cite{qsym} utilizes instruction-level selective symbolic execution to make hybrid fuzzing scalable enough to test large applications.
Although there have been many methods to optimize the efficiency of hybrid fuzzing, it still faces the problem of being time-consuming when complex relations exist between path constraints. 
In contrast, \tool{} passes path constraints in a lightweight manner to speed up the exploration of hard-to-cover branches.

\noindent\textbf{Taint-based Fuzzing.}
Taint analysis builds the relationships between input bytes and path constraints.
Taint-based fuzzers use taint tracking to determine which part of the input will be mutated to reach the target code blocks. Angora~\cite{angora} utilizes the byte-level taint analysis to track which input bytes flow into path constraints. Then Angora mutates only these bytes based on a gradient descent algorithm to solve path constraints. 
To explore deeply nested conditional statements, Matryoshka~\cite{matryoshka} identifies nesting conditional statements and mutates the input to solve all path constraints simultaneously. VUzzer~\cite{vuzzer} tracks the related bytes of comparisons of magic bytes and generates inputs to satisfy the constraints. 

While taint analysis is efficient in resolving constraints on hard-to-cover branches, it can also be time-consuming. Thus, some methods propose more lightweight methods to help track such taint information. Steelix~\cite{steelix} utilizes comparison progress information to penetrate magic bytes more effectively.
REDQUEEN~\cite{redqueen} exploits input-to-state correspondence to solve hard fuzzing problems. 
ProFuzzer~\cite{profuzzer} iterates through all possible values and collects the corresponding execution profiles for each byte. Then ProFuzzer groups fields and determines their corresponding type from these profiles. 
WEIZZ~\cite{weizz} flips one bit at a time on the entire input and checks for changes in comparison operands to infer potential dependency between bits and branch conditions.
GREYONE~\cite{greyone} mutates input bytes in a pilot stage and monitors the variable value changes to infer the tainted bytes.
To make taint analysis path-aware, PATA~\cite{pata} records values of each occurrence of variables. 

While ensuring accuracy, inference-based fuzzers are simple yet efficient. However, the relationship between their efficiency and input length is strong. Since the pilot stage mutates inputs byte by byte, inference can become time-consuming when analyzing long inputs. 
On the contrary, \tool{} formulates byte selection as Shapley analysis and does not analyze each byte additionally.
Thus, The efficiency of \tool{} is independent of the seed length.
On the other hand, taint-based fuzzers and inference-based fuzzers focus on a single path constraint and neglect the relations between path constraints. 
\tool{} assigns more energy to bytes with higher Shapley values to slove the complex path constraints.





\noindent\textbf{Other Fuzzing Strategies.}
NEUZZ~\cite{neuzz} builds the relationships between input bytes and branch behaviors via deep learning models. Then NEUZZ preferentially mutates the bytes with high score, indicating those bytes have high chance to be related to branches.
As the same program logic can be repeatedly used within a single program, EMS~\cite{ems} captures the byte-level mutation strategies from intra- and inter-trial history to explore new paths and crashes.
While EMS focuses on the values of input segments, our \tool{} focuses on the bytes of inputs. 
A byte may be used as a path constraint in different program logic, and the different program logic often uses different CMP instructions.

\section{Conclusion}
In this paper, we identify that the repeated mutation of a small portion of positions contributes to the most of edge discovery based on the results of Shapley analysis. 
Therefore, the bytes with high Shapley values deserve more computational resources and can be used to accelerate the discovery of new edges.
Motivated by this insight, we propose \tool{}, a novel fuzzer that aims to increase code coverage with a Shapley-guided byte schedule.
We formulate byte mutation as Shapley analysis, and gather seeds into families based on the formulation. To improve the performance of Shapley-guided mutation, \tool{} utilizes a contextual multi-armed bandit approach to optimize the use of Shapley values.
Based on the experiments on two third-party testing platforms, UNIFUZZ and MAGMA, we demonstrate that \tool{} outperforms several state-of-the-art fuzzers, such as NEUZZ, Greyone and EMS, in both code coverage and bug discovery.

\section*{Acknowledgment}
Xi Xiao is supported in part by the Overseas Research Cooperation Fund of Tsinghua Shenzhen International Graduate School (HW2021013).
\bibliographystyle{IEEEtranS}
\bibliography{reference}


\end{document}